\documentclass[aps, pra, superscriptaddress, notitlepage, reprint]{revtex4-1}
\usepackage[english]{babel}
\usepackage{amsmath,amsthm}
\usepackage{amsfonts}
\usepackage[pdfborder={0 0 0}, colorlinks=true, urlcolor=blue, linkcolor=blue, citecolor=blue]{hyperref}
\usepackage{color}
\usepackage{graphicx}
\usepackage{float}
\usepackage{subfigure}
\usepackage{lipsum}

\theoremstyle{definition}

\theoremstyle{remark}

\begin{document}

\title{Entanglement and dynamical phase transition in a spin-orbit-coupled Bose-Einstein condensate}
\author{F. X. Sun}
\affiliation{State Key Laboratory of Mesoscopic Physics, School of Physics, Peking University, Collaborative Innovation Center of Quantum Matter, Beijing 100871, China}
\affiliation{Collaborative Innovation Center of Extreme Optics, Shanxi University, Taiyuan 030006, China}
\author{W. Zhang}
\email{wzhangl@ruc.edu.cn}
\affiliation{Department of Physics, Renmin University of China, Beijing 100872, China}
\affiliation{Beijing Key Laboratory of Opto-Electronic Functional Materials and Micro-Nano Devices, Renmin University of China, Beijing 100872, China}
\author{Q. Y. He}
\email{qiongyihe@pku.edu.cn}
\affiliation{State Key Laboratory of Mesoscopic Physics, School of Physics, Peking University, Collaborative Innovation Center of Quantum Matter, Beijing 100871, China}
\affiliation{Collaborative Innovation Center of Extreme Optics, Shanxi University, Taiyuan 030006, China}
\author{Q. H. Gong}
\affiliation{State Key Laboratory of Mesoscopic Physics, School of Physics, Peking University, Collaborative Innovation Center of Quantum Matter, Beijing 100871, China}
\affiliation{Collaborative Innovation Center of Extreme Optics, Shanxi University, Taiyuan 030006, China}
\begin{abstract}

Characterizing quantum phase transitions through quantum correlations has been deeply developed for a long time, while the connections between dynamical phase transitions (DPTs) and quantum entanglement is not yet well understood. In this work, we show that the time-averaged two-mode entanglement in the spin space reaches a maximal value when it undergoes a DPT induced by external perturbation in a spin-orbit-coupled Bose-Einstein condensate. We employ the von Neumann entropy and a correlation-based entanglement criterion as entanglement measures and find that both of them can infer the existence of DPT. While the von Neumann entropy works only for a pure state at zero temperature and requires state tomography to reconstruct, the experimentally more feasible correlation-based entanglement criterion acts as an excellent proxy for entropic entanglement and can determine the existence of entanglement for a mixed state at finite temperature, making itself an excellent indicator for DPT. Our work provides a deeper understanding about the connection between DPTs and quantum entanglement, and may allow the detection of DPT via entanglement become accessible as the examined criterion is suitable for measuring entanglement.

\end{abstract}
\maketitle

\section{Introduction}

As one of the most intriguing features of quantum mechanics, quantum entanglement has been regarded as a key resource to detect and understand properties of complex many-body systems. For instance, recently a great deal of effort has been devoted to establishing a deep understanding about quantum phase transitions (QPTs) via the examination of their entanglement behaviors~\cite{osterloh2002scaling,vidal2003entanglement,lambert2004entanglement,gu2004entanglement,le2008entanglement,pereira2016effects,wu2004quantum,wu2006linking,yang2005reexamination,braun2017quantum,vidal2006concurrence}. Quantum phase transition is defined as a transition between distinct ground states of quantum many-body systems when a controlled parameter in the Hamiltonian crosses a critical point~\cite{sachdev2011quantum}. Compared with the great development of studying QPTs in equilibrium systems, the understanding of the non-equilibrium dynamical phase transitions (DPTs) is still inadequate~\cite{polkovnikov2011colloquium}. Although some investigations have been done to study the properties of DPTs~\cite{lo1990ising,jung1990scaling,ryu1996dynamical,sides1998kinetic,yuzbashyan2006relaxation,sciolla2010quantum,diehl2010dynamical,calabrese2011quantum,mitra2012time,heyl2013dynamical,heyl2014dynamical,klinder2015dynamical}, only few works linked it with entanglement~\cite{lin2016non,jurcevic2017direct,Zhang2017observation}. Recently, the observation of many-body DPTs with up to 10 trapped ion qubits has shown that DPTs in the simulated Ising models can control entanglement production~\cite{jurcevic2017direct}. Another elegant experiment with a quantum simulator composed of up to 53 ion qubits with long-range Ising interactions has also uncovered the connection between DPTs and many-body correlations~\cite{Zhang2017observation}. Inspired by those achievements, we investigate theoretically an extension of this connection in quantum systems other than Ising model. For instance, a rich variety of quantum phases in spin-orbit-coupled (SOC) Bose-Einstein condensates (BECs) has been investigated theoretically and experimentally~\cite{dalibard2011colloquium, galitski2013spin, zhou2013unconventional, goldman2014light, zhai2015degenerate, zhang2016properties}. A proposal of simulating spin DPT by ultra-cold atoms has been reported in this system~\cite{poon2016quantum}. We are wondering whether this kind of DPT can be characterized by the behavior of two-mode entanglement in the synthetic spin space. 

In this paper, by introducing an external perturbation (switching on an additional lattice potential) in the Hamiltonian of a spin-1/2 BEC of $^{87}$Rb atoms with 1D synthetic spin-orbit coupling, we study the behavior of dynamical two-mode entanglement in spin space and use it to characterize spin DPT. A previous analysis~\cite{poon2016quantum} shows that the additional lattice potential can drive a periodic evolution of the system in spin space, and there exists a DPT between magnetized and unmagnetized states at a critical lattice depth. By examining the entropic entanglement measure~\cite{hines2003entanglement, xie2006quantum, byrnes2012quantum}, we show that the periodic motion in spin space leads to the periodic evolution of two-mode entanglement, and the time-averaged entropic entanglement over an oscillation period reaches a maximal value at the DPT. 

Although the entropy of entanglement provides an excellent indicator of DPT in such a system, it works only for pure states at zero temperature and requires reconstruction of the quantum states via tomography in experiments. To overcome these difficulties, we also use a correlation-based entanglement criterion that is suitable for measuring to detect DPT and study the influence of thermal excitations. Specifically, a criterion was originally introduced by Hillery and Zubairy (HZ)~\cite{hillery2006entanglement} and developed for double-well BEC systems~\cite{he2011einstein, he2012einstein}.  We show that the HZ criterion is an excellent proxy for entropic entanglement measure, hence can be used to characterize the DPT. Moreover, we find that the thermal effects will change the critical point of the DPT, which can also be confirmed by the shifts of the maximum of time-averaged HZ entanglement parameter. 

The remainder of this paper is organized as follows. In Sec.~\ref{Sec:Model}, we introduce the model of DPT induced by external perturbation in BEC with SOC. We then characterize the two-mode entanglement of the system across the DPT via von Neumann entropy and HZ criterion in Sec.~\ref{Sec:entropy} and Sec.~\ref{Sec:HZ}, respectively. The results show that the behavior of entanglement can infer the existence of DPT. In Sec.~\ref{Sec:thermal}, we discuss finite temperature effect and show that the entanglement is robust to thermal excitations and can also be used to signature DPT. Finally we summarize our results and discuss the experimental feasibility in Sec.~\ref{Sec:conclusion}.

\section{Model}\label{Sec:Model}

We consider a two-component BEC of $^{87}$Rb atoms with one-dimensional (1D) synthetic spin-orbit coupling. As discussed in Ref.~\cite{poon2016quantum}, such a system features a quantum spin DPT in the presence of an additional lattice potential as external perturbation. By labeling the two atomic components as (pseudo-)spin up and down, the Hamiltonian reads (with natural units $\hbar=m=1$)
\begin{eqnarray}
	H&=&H_{0}+H_{\rm int} ,\nonumber\\
	H_{0}&=&\sum_{s,s'=\uparrow,\downarrow}\int d^3r \psi_s^\dagger \left( -\frac{\nabla_r^2}{2}+ik_0\partial_x\sigma_z+\frac{\Omega}{2}\sigma_x \right)_{ss'}\psi_{s'} , \nonumber\\
	H_{\rm int}&=&\int d^3r \frac{g_s}{2}\left(\psi_{\downarrow}^{\dagger}\psi_{\downarrow}^{\dagger}\psi_{\downarrow}\psi_{\downarrow}+\psi_{\uparrow}^{\dagger}\psi_{\uparrow}^{\dagger}\psi_{\uparrow}\psi_{\uparrow}\right) \nonumber\\
	&&+\int d^3r g_a \psi_{\downarrow}^{\dagger}\psi_{\uparrow}^{\dagger}\psi_{\uparrow}\psi_{\downarrow}.
\end{eqnarray}
Here $H_0$ is the single-particle Hamiltonian including 1D SOC along the $x$ direction, and $H_{\rm int}$ is the interaction term. The field operators $\psi_{s}$ ($\psi_{s}^{\dagger}$) annihilates (creates) an atom with spin $s=\uparrow, \downarrow$, $\Omega$ is the Raman coupling strength, and $k_0$ is determined by the Raman laser's wave vector. Note that we assume a spin-symmetric interaction with $g_{\uparrow\uparrow}=g_{\downarrow\downarrow}=g_{s}$ and $g_{\uparrow\downarrow}=g_{a}$, and the two-photon detuning $\delta=0$ for simplicity. 

By diagonalizing $H_{0}$, the eigenenergies of two subbands are given by $E_{k}^{\pm}={k^2}/{2}\pm \sqrt{k_0^2k_x^2+{\Omega^2}/{4}}$. For $\Omega<2k_0^2$, the lower subband has a double-minimum structure at $k_x=\pm k_{\rm min}$ with $k_{\rm min}=k_0\sqrt{1-\Omega^2/4k_0^4}$. For $\Omega>2k_0^2$, it has only a single minimum at $k_{\rm min}=0$. This structure is the origin of phase transitions among magnetized plane-wave phase, unpolarized stripe phase, and zero-momentum normal phase.

When the interaction is concerned, we can assume that the ground-state wave function of the condensate takes the form
\begin{equation}
	\Psi=\sqrt{n}\left[ \alpha\left( 
			\begin{array}{c}
				\cos\theta \\ -\sin\theta
			\end{array}
	\right)e^{ik_mx}+\beta\left( 
		\begin{array}{c}
			\sin\theta \\ -\cos\theta
		\end{array}
	\right)e^{-ik_mx} \right] ,
	\label{ground-state}
\end{equation}
where $\alpha$ and $\beta$ are arbitrary complex numbers satisfying $|\alpha|^2+|\beta|^2=1$, and $\tan2\theta=\Omega/(2k_0k_m)$. In general, the momenta of many-body eigenstates $\pm k_m$ are different from the minimal positions of single-particle dispersion $\pm k_{\rm min}$. But in our parameter region where $\Omega/k_0^2<0.6$ and $g_s n \approx 1.0k_0^2$ ($n=n_{\uparrow}+n_\downarrow$ represents the condensate density), we can safely set $k_m\simeq k_{\rm min}$. The remarkable property is that there exist two critical Raman couplings~\cite{li2012quantum}, given by
\begin{eqnarray}
	\Omega_{c1}&=&2(k_0^2-2G_2) ,\\
	\Omega_{c2}&=&2\sqrt{(k_0^2+G_1)(k_0^2-2G_2)\frac{2G_2}{G_1+2G_2}} ,
\end{eqnarray}
where $\Omega_{c1}>\Omega_{c2}$ and $G_{1}={n}(g_{s}+g_{a})/4$, $G_{2}={n}(g_{s}-g_{a})/4$. When $\Omega<\Omega_{c2}$, the ground state is a superposition of states with $k_x=\pm k_{m}$ ($|\alpha|^2=|\beta|^2=1/2$), which is referred as the stripe phase. For $\Omega_{c2}<\Omega<\Omega_{c1}$, there exist two degenerate states with $k_x=k_m$ ($|\alpha|=1$, $|\beta|=0$) or $k_x=-k_m$ ($|\alpha|=0$, $|\beta|=1$), called the magnetized phase. If the Raman coupling is large enough that $\Omega>\Omega_{c1}$, the ground state is at $k_x=0$, giving the normal phase~\cite{dalibard2011colloquium, galitski2013spin, zhou2013unconventional, goldman2014light, zhai2015degenerate, zhang2016properties, li2012quantum, martone2012anisotropic}.

\begin{figure}[t]
	\centering
	\includegraphics[width=0.35\textwidth]{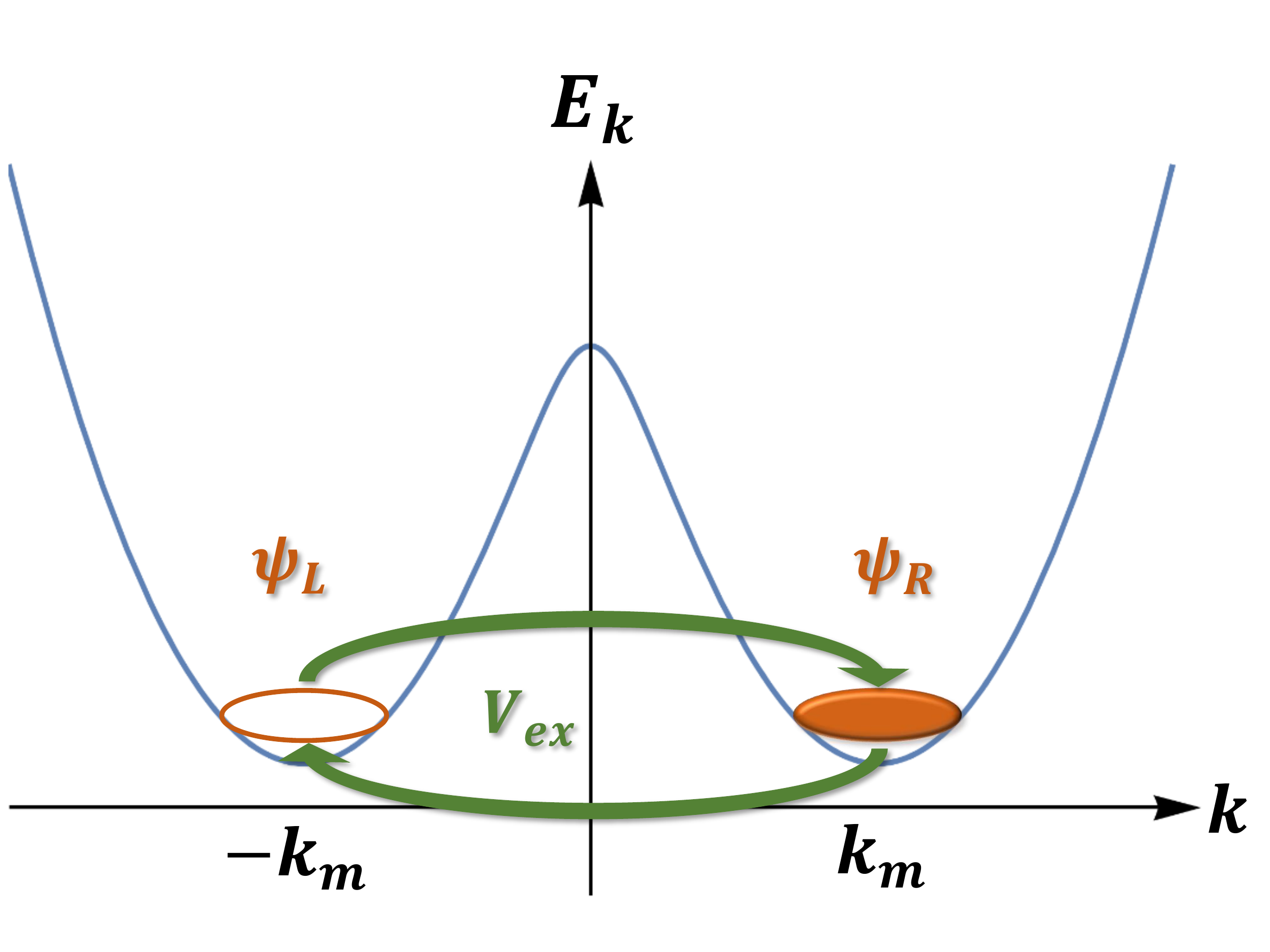}
	\caption{The external perturbation $V_{\rm ex}$ described by Eq.~(\ref{Vex}) can induce resonant couplings between the two magnetized phases, marked as $\psi_{R}$ and $\psi_{L}$ at $k_m$ and $-k_m$, respectively. }
	\label{fig:BEC_ground_state}
\end{figure}

For a magnetized phase with, e.g., $k_x=k_m$ ($|\alpha|=1, |\beta|=0$), an additional external potential is switched on at $t=0$~\cite{poon2016quantum},
\begin{equation}
	V_{\rm ex}(r,t)=\left\{
		\begin{array}{lc}
			0, & t<0,\\
			V_{0}\int d^3r \cos^2k_m x(\psi_{\uparrow}^{\dagger}\psi_{\uparrow}+\psi_{\downarrow}^{\dagger}\psi_{\downarrow}), & t>0,
		\end{array}
	\right. 
	\label{Vex}
\end{equation}
where $V_0$ is the strength of the perturbation. $V_{\rm ex}$ can drive resonant couplings between the two degenerate magnetized phases $\psi_R$ and $\psi_L$ at $k_{m}$ and $-k_m$, respectively, as indicated schematically in Fig.~\ref{fig:BEC_ground_state}. The normalized time-dependent condensate wave function takes the form 
\begin{equation}
	|\Psi_{\rm BEC}(r,t)\rangle = \frac{1}{\sqrt{N!}} \left[ \alpha^{\ast}(t)\psi_{R}^{\dagger}+\beta^{\ast}(t)\psi_{L}^{\dagger} \right]^{N}|vac\rangle ,
	\label{wave-function}
\end{equation}
where $N$ is the total number of atoms, $|vac\rangle$ denotes the vacuum state, and $\psi_{R}=(\cos\theta\psi_{\uparrow}-\sin\theta\psi_{\downarrow})e^{ik_mx}$, $\psi_{L}=(\sin\theta\psi_{\uparrow}-\cos\theta\psi_{\downarrow})e^{-ik_mx}$ are the field operators.

The quantum spin dynamics of the system under the perturbation given in Eq.~(\ref{Vex}) has been studied in Ref.~\cite{poon2016quantum}. It is found that a critical external perturbation strength
\begin{equation}
	V_{0,{\rm crit}}=\frac{2(E_{s}-E_{m})}{\sin2\theta} 
	\label{V0c}
\end{equation}
is required to have a full transition from one magnetized phase to another. Specifically, when the perturbation strength is below $V_{0,{\rm crit}}$, the maximum number of atoms that can transit from the initial state with $k_x=k_m$ to the other degenerate state at $k_x=-k_m$ is less than half of the total atom number. On the other hand, when the perturbation strength $V_{0}>V_{0,{\rm crit}}$, all atoms can fully transit during time evolution. Thus, the critical perturbation strength $V_{0,{\rm crit}}$ reveals a quantum DPT. We also stress that the critical point $V_{0,{\rm crit}}$ depends on the condensate interaction energy $g_sn$ via $E_{s}=2G_{1}\cos^{2}\theta\sin^{2}\theta$ and $E_{m}=G_{2}\cos^{2}2\theta$, which characterize the interaction energies for the stripe phase and magnetized phase, respectively. 

The transition mentioned above can be characterized by an order parameter defined as the time average of spin polarization $\langle s_z(t)\rangle= |\alpha(t)|^2-|\beta(t)|^2$ ($s_z$ is the Pauli matrix acting on the pseudospin space spanned by the two magnetized states) over an oscillation period $T_R$~\cite{poon2016quantum}
\begin{equation}
	\bar{M}=\frac{1}{T_{R}}\int_{0}^{T_{R}}\langle s_{z}(t)\rangle dt .
	\label{order-parameter}
\end{equation}
It is straightforward to check that $\bar{M}> 0$ when the perturbation strength is lower than the critical point and the system is in the dynamical magnetized phase, and $\bar{M}\equiv 0$ when the system is dynamically non-magnetized with perturbation strength exceeding the critical point.

\section{Entropy of entanglement}\label{Sec:entropy}

Previous works have studied entanglement of pure states of bipartite systems using entropy of entanglement, which is the von Neumann entropy of the reduced density operator of either of the subsystems. In this section we first study whether the quantum spin DPT at zero temperature can be characterized by the entropic entanglement measure. As the two degenerate ground states of the magnetized phase (labeled by $\psi_{R}$ and $\psi_{L}$) can be considered as an analogue of a double-well BEC in momentum space, we can adopt the same method which is commonly used in spatially separated double-well BEC~\cite{spekkens1999spatial,he2012einstein,he2011einstein,vidal2007entanglement}, and treat $\psi_{R}$ and $\psi_{L}$ as two modes within the two-mode approximation~\cite{spekkens1999spatial}. Our results show that the behavior of the entropic entanglement can act as a indicator of the quantum spin DPT in this system.

By applying the two-mode approximation, the time-dependent wave function in Eq.~(\ref{wave-function}) can be expanded in term of Fock states as
\begin{equation}
	|\Psi_{\rm BEC}(t)\rangle = \sum_{n=0}^{N} \sqrt{C_{N}^{n}} \alpha^{\ast n}\beta^{\ast N-n}|n,N-n\rangle_{R,L} ,
	\label{two-mode_wave_function}
\end{equation} 
where $C_{N}^{n}$ is the binomial coefficient, and the Fock basis are defined as
\begin{equation}
	|n,N-n\rangle_{R,L}=\frac{(\psi_{R}^{\dagger})^{n}}{\sqrt{n!}}\frac{(\psi_{L}^{\dagger})^{N-n}}{\sqrt{(N-n)!}}|vac\rangle .
	\label{number_state}
\end{equation}

The dynamics can be described by defining the field operator $\psi(t)=\alpha(t)\psi_{R}+\beta(t)\psi_{L}$, and considering Heisenberg equation
\begin{equation}
	i\frac{d\psi(t)}{dt}=\left[ \psi(t), H_{0}+H_{\rm int}+V_{\rm ex} \right],
	\label{Heisenberg}
\end{equation}
leading to the following equation of motion
\begin{equation}
	i\frac{d}{dt}\left( 
		\begin{array}{c}
			 \alpha(t) \\ \beta(t)
		\end{array}
\right)=H_{\rm eff}\left( 
	\begin{array}{c}
		\alpha(t) \\ \beta(t)
	\end{array}
\right).
	\label{motion}
\end{equation}
Here, the effective two-mode Hamiltonian is given by
\begin{eqnarray}
	H_{\rm eff}&=&E_{k}^{-}+\frac{V_{0}}{2}+G_{1}+V_{p}s_{x}+E_{m}(|\alpha|^{2}-|\beta|^{2})s_{z} \nonumber\\
	&&+2E_{s}\left[ {\rm Re}(\alpha\beta^{\ast})s_{x}-{\rm Im}(\alpha\beta^{\ast})s_{y} \right],
	\label{Heff}
\end{eqnarray}
where $s_{x,y,z}$ are spin matrices spanned by the two magnetized states, and $V_{p}=V_{0}\cos\theta\sin\theta/2$ represents the coupling strength induced by external perturbation.

\begin{figure*}[t]
	\centering
	\includegraphics[width=0.245\textwidth]{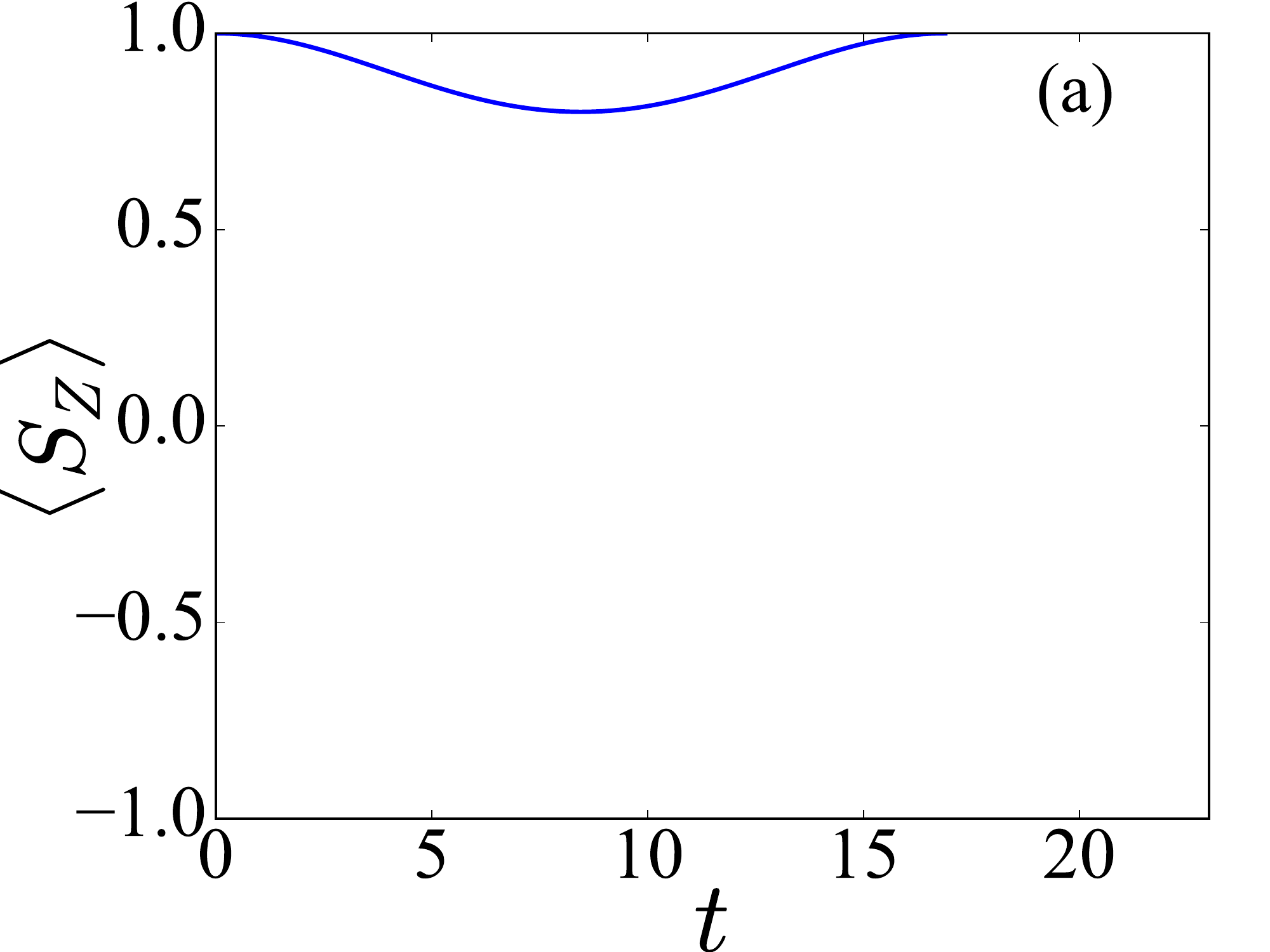}
	\includegraphics[width=0.245\textwidth]{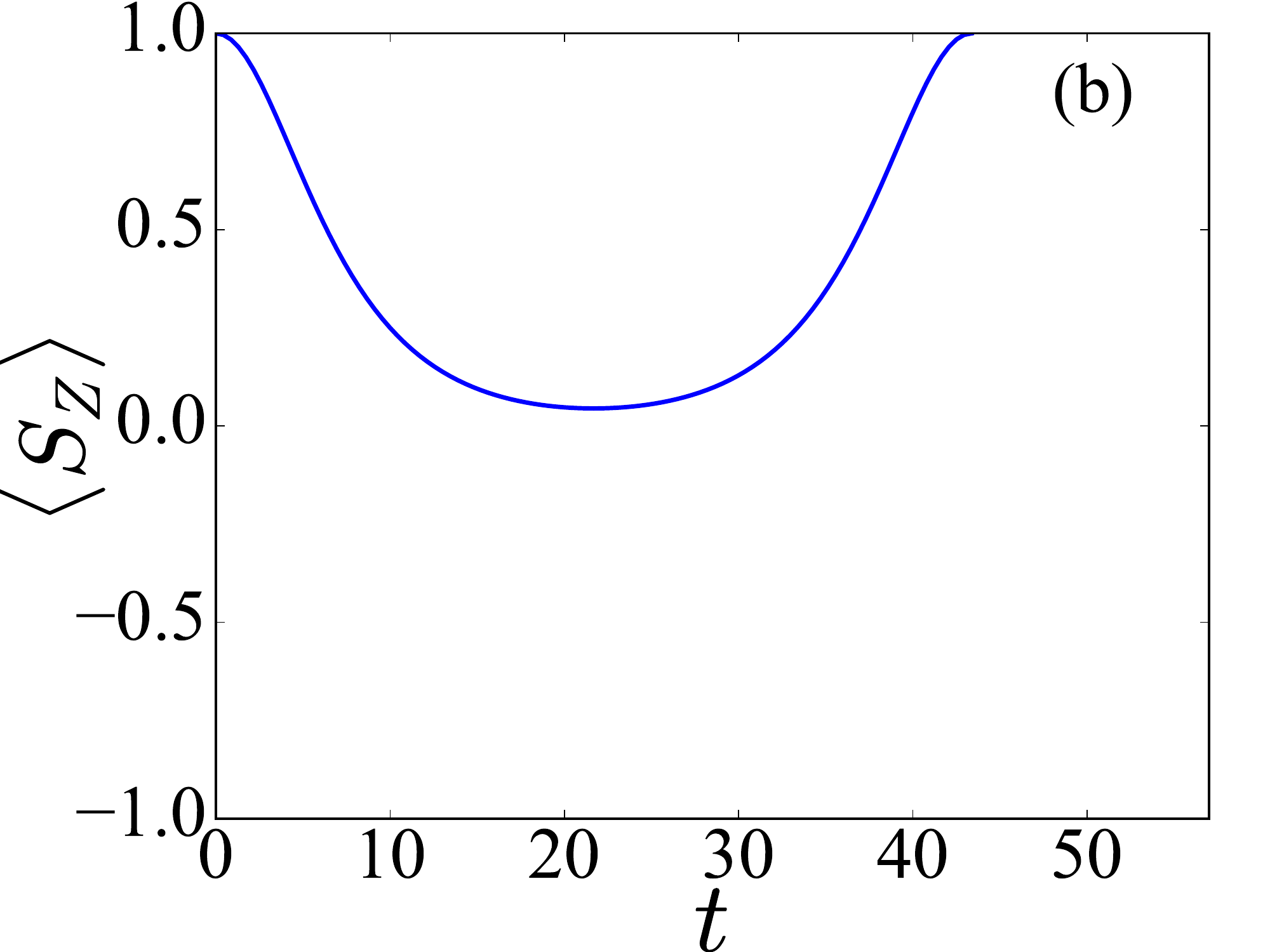}
	\includegraphics[width=0.245\textwidth]{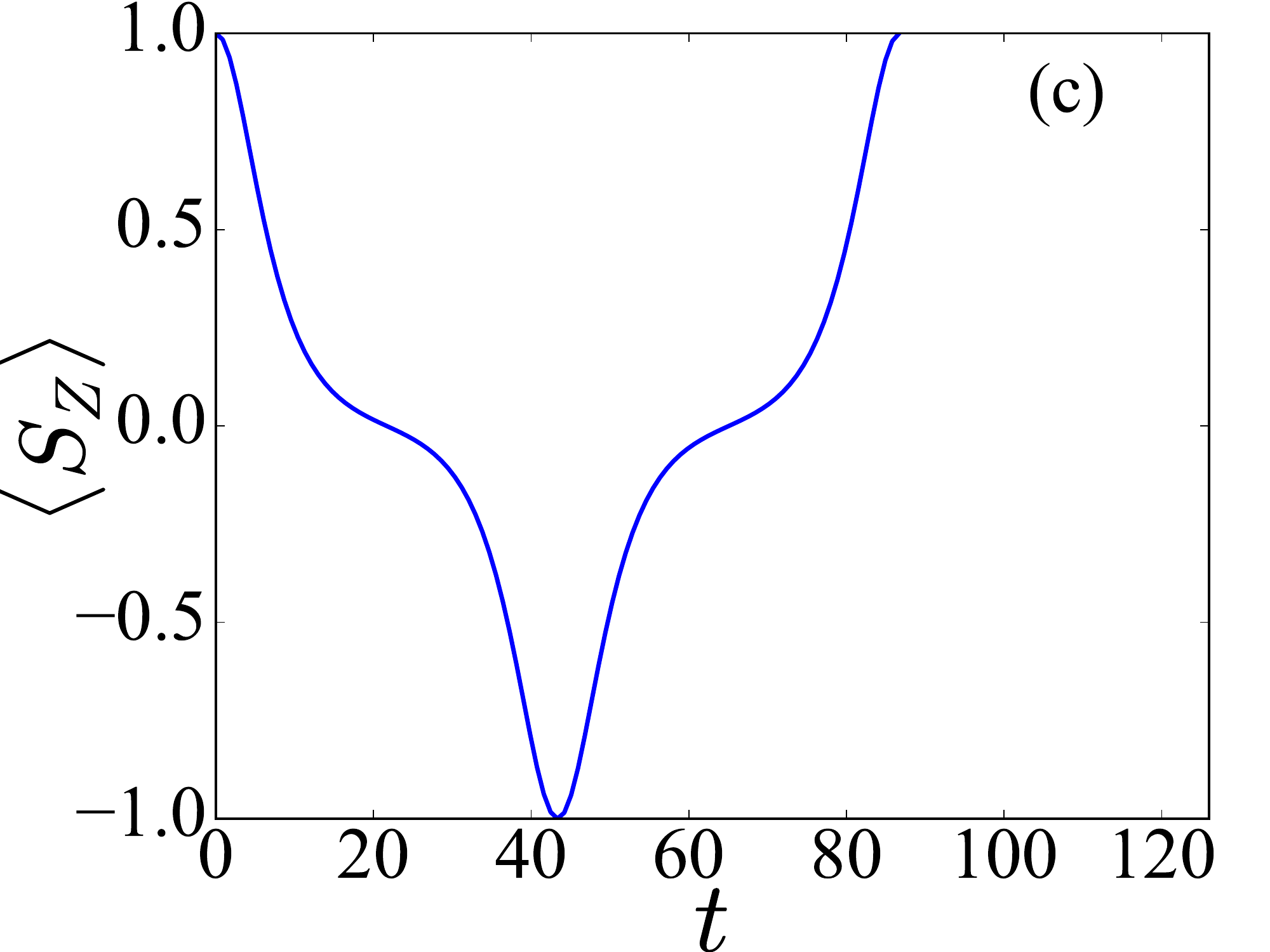}
	\includegraphics[width=0.245\textwidth]{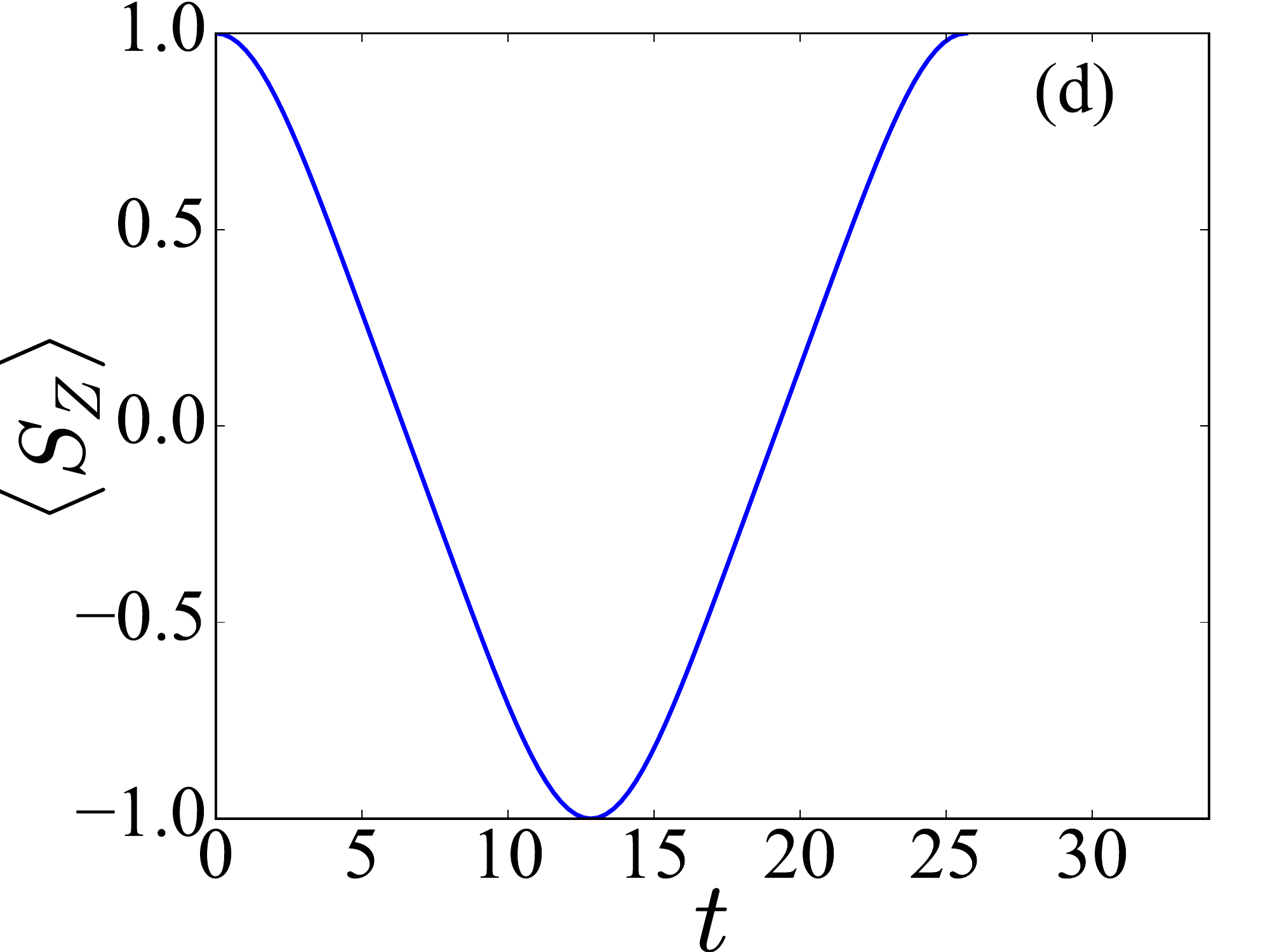}\\
	\includegraphics[width=0.245\textwidth]{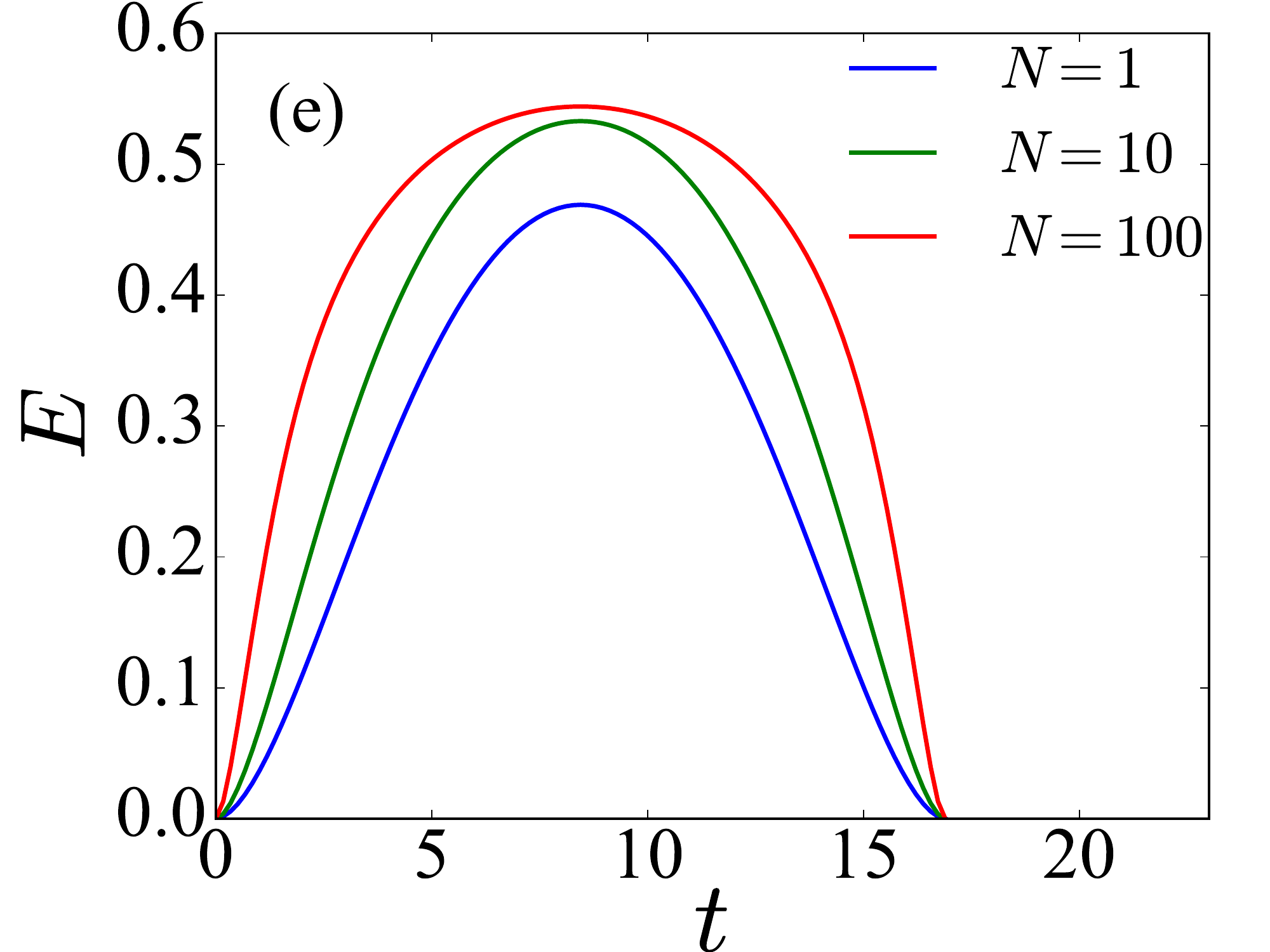}
	\includegraphics[width=0.245\textwidth]{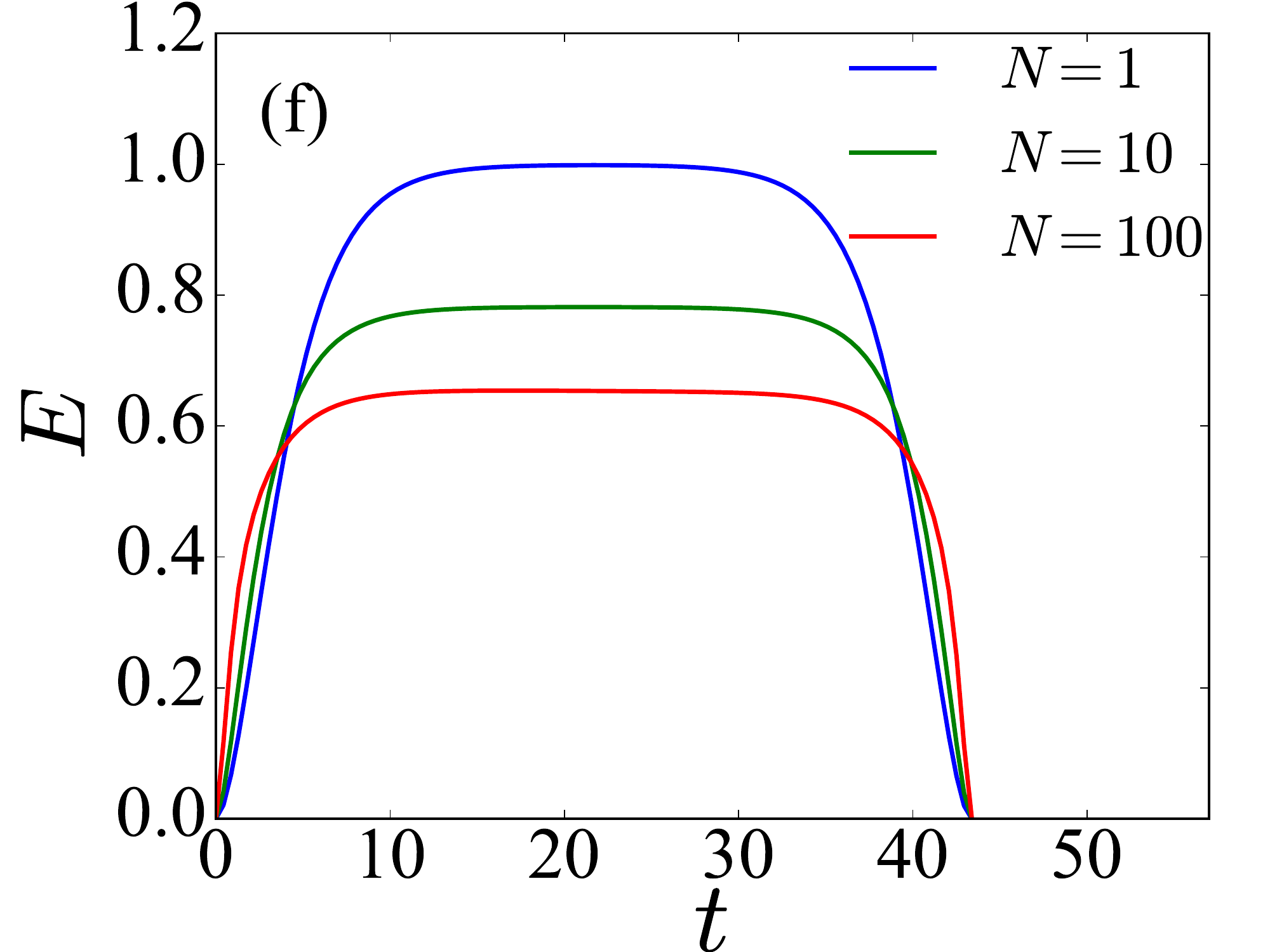}
	\includegraphics[width=0.245\textwidth]{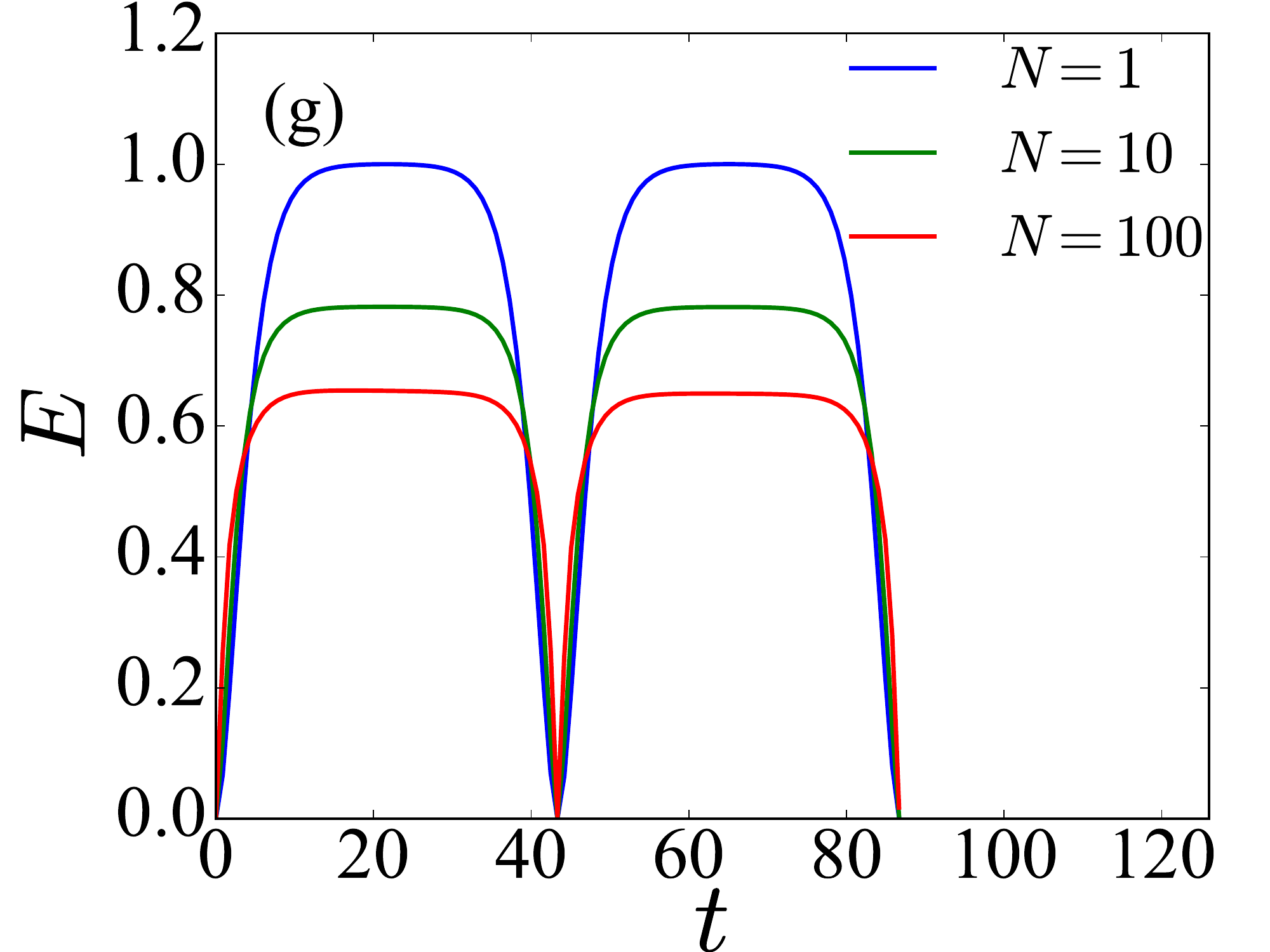}
	\includegraphics[width=0.245\textwidth]{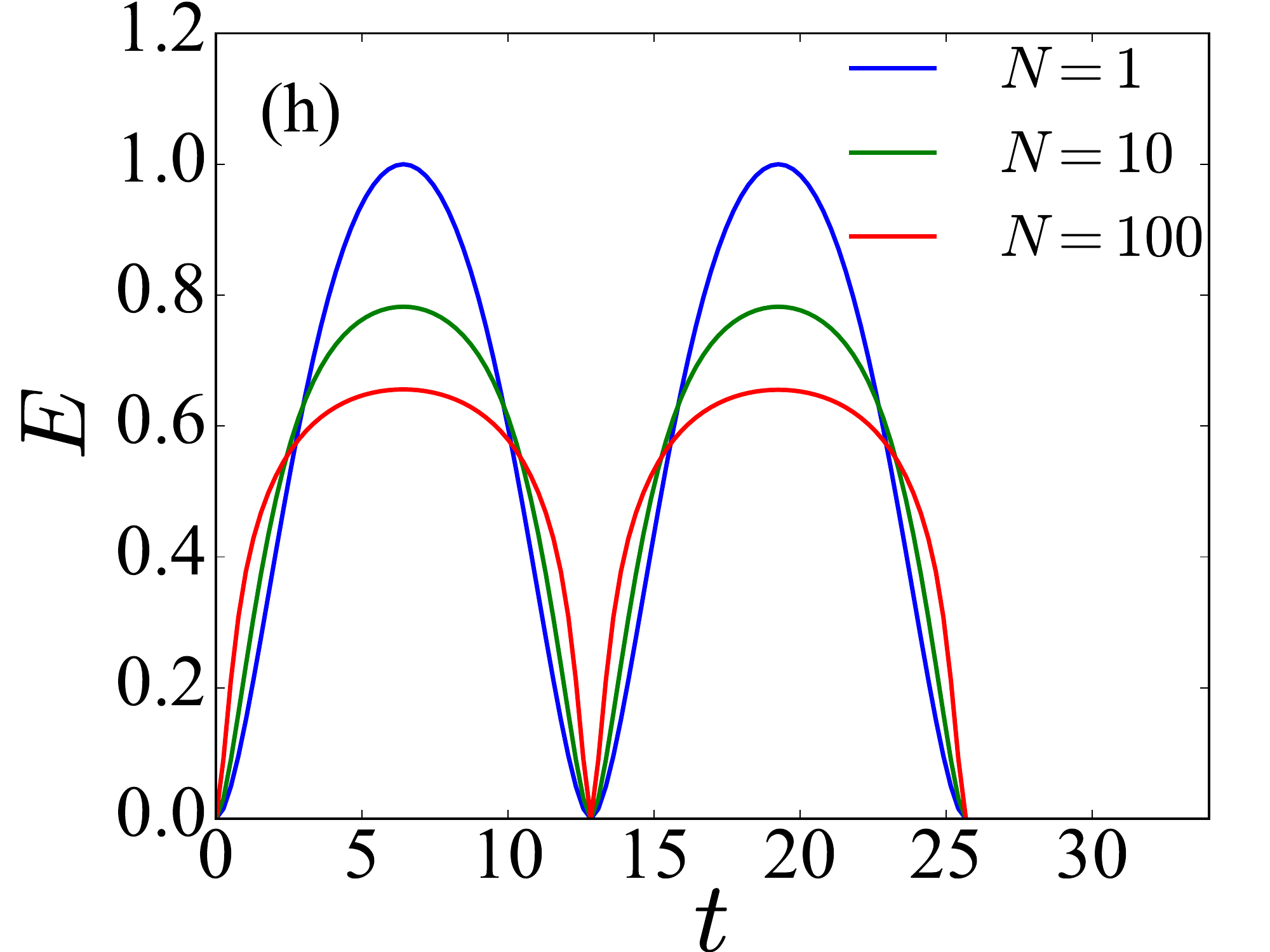}
\caption{The evolution of entropic entanglement $E$ is related to the evolution of pseudospin $\langle s_z\rangle$. From left to right, the external perturbation strength is (a)(e) $V_0 = 0.6V_{0,{\rm crit}}$, (b)(f) $0.999V_{0,{\rm crit}}$, (c)(g) $1.001V_{0,{\rm crit}}$, and (d)(h) $1.4V_{0,{\rm crit}}$. For each situation, we plot the entropic entanglement for various numbers of atoms $N=1$ (blue), $N=10$ (green), and $N=100$ (red solid). Other parameters are used as $g_{s}n=1.0k_{0}^{2}$ and $\Omega=0.3k_{0}^{2}$.}
\label{fig:entropy}
\end{figure*}

Theoretically, the von Neumann entropy $E_{\rm vn}=-{\rm Tr}\left[ \rho_{R}\log\rho_{R} \right]$ can be used to evaluate the entanglement between two subsystems, where $\rho_{R}={\rm Tr}_{L}[\rho_{R,L}]$ is the reduced density operator for $\psi_R$ and $\rho_{R,L}=|\Psi_{\rm BEC}\rangle\langle\Psi_{\rm BEC}|$. Considering the specific form of the time-dependent wave function $|\Psi_{\rm BEC}\rangle$ described in Eq.~(\ref{two-mode_wave_function}), the entropy of entanglement between $\psi_R$ and $\psi_L$ thus reads
\begin{equation}
	E_{\rm vn}=-\sum_{n=0}^{N}C_{N}^{n}|\alpha|^{2n}|\beta|^{2(N-n)}\log_2
	\left[C_{N}^{n}|\alpha|^{2n}|\beta|^{2(N-n)}\right].
	\label{entanglement_entropy}
\end{equation}
Here, $E_{\rm vn}=0$ for separable product states, and $E_{\rm max}=\log_2(N+1)$ for maximally entangled states when all atoms are equally represented~\cite{hines2003entanglement}. As a measure of the entanglement, we plot in Fig.~\ref{fig:entropy} the ratio of $E_{\rm vn}$ to its corresponding maximum value $E=E_{\rm vn}/E_{\rm max}$ for various numbers $N$ and external perturbation strength $V_0$~\cite{hines2003entanglement, xie2006quantum, byrnes2012quantum}. The values of $E$ range from $0$ to $1$. Note that the ratio $E$ only represents how much the entanglement of the state $|\Psi_{\rm BEC}\rangle$ with total $N$ atoms is less than its corresponding maximum entanglement, $\log(N+1)$. It is meaningless to compare the values of $E$ for different $N$. As a comparison, the dynamical evolution of the spin polarization $\langle s_z \rangle$ is also shown using the same parameter $V_0$. 

From Fig.~\ref{fig:entropy}, it is clear that the entropic entanglement evolves with time periodically, and the period is related to that of the spin oscillation. In the regime where the perturbation strength is below the critical value $V_{0}<V_{0,{\rm crit}}$, as shown in Figs.~\ref{fig:entropy}(a)(b)(e)(f), only less than half of the atoms can transit from one magnetized state at $k_x=k_m$ to the other at $k_x=-k_m$. Thus the spin evolves only on the upper half spherical surface of the Bloch sphere, i.e., $\langle s_z\rangle>0$. In such a case, the oscillatory period of the two-mode entanglement is as same as that of the spin oscillation. When all the atoms are at the magnetized state $k_x=\pm k_m$, we have $\langle s_z\rangle=\pm1$ and $E=0$, i.e., the two-mode entanglement doesn't exist. When atoms distribute equally in the two modes, i.e., $\langle s_{z}\rangle$ approaches zero, the two-mode entanglement reaches a maximal value. On the other hand, in the regime where the external perturbation strength exceeds the critical value, as shown in Figs.~\ref{fig:entropy}(c)(d)(g)(h), the spin evolves on the entire spherical surface of the Bloch sphere for one oscillation period $T_R$. A complete transition of atoms from one magnetized phase to another occurs, i.e., $\langle s_z\rangle=-1$, at $T_R/2$, where $E=0$ ends one complete cycle of entanglement motion. Therefore, for this case the period of entanglement oscillation is half of that of the spin evolution. 
\begin{figure}[t]
	\centering
	\includegraphics[width=0.4\textwidth]{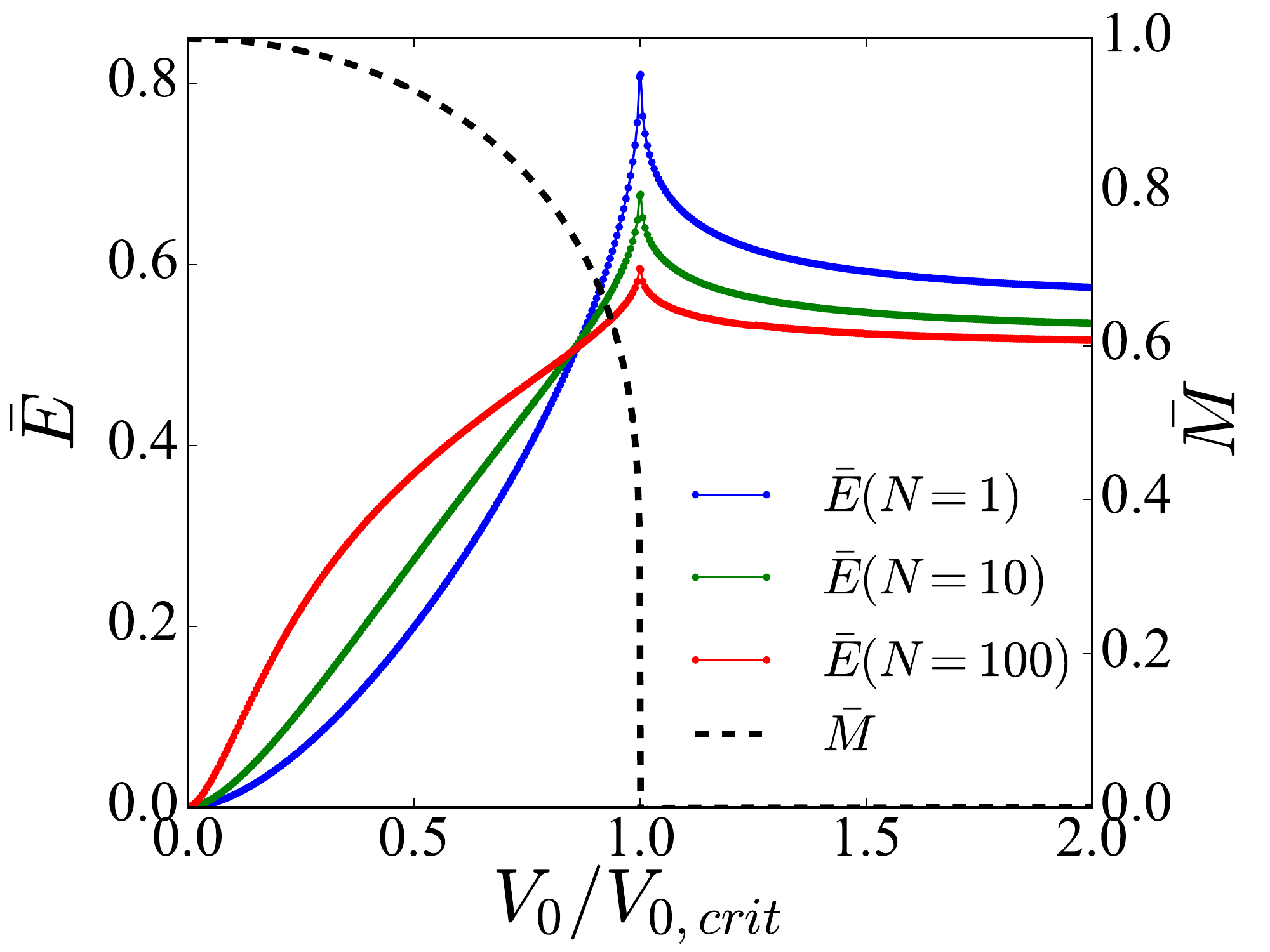}
	\caption{The time-averaged entropic entanglement measure $\bar{E}$ calculated as a function of perturbation strength $V_0/V_{0,{\rm crit}}$ (left axis). Different lines represent different total numbers of atoms $N=1$ (blue), $N=10$ (green), and $N=100$ (red). The entanglement displays a sharp peak at the critical value $V_{0,{\rm crit}}$ where the DPT occurs as characterized by the order parameter $\bar{M}$ (black dashed, right axis). This observation suggests that the critical behavior of entropic entanglement measure can be used to identify the DPT. Other parameters are used as those in Fig.~\ref{fig:entropy}.}
	\label{fig:time-average_entropy}
\end{figure}

The spin DPT across $V_{0,{\rm crit}}$ can be characterized by the order parameter $\bar{M}$ defined in Eq.~(\ref{order-parameter}), which is depicted by the dashed black line in Fig.~\ref{fig:time-average_entropy}. To reveal the connection between DPT and entanglement, we define the time-averaged entropic entanglement measure as
\begin{equation}
	\bar{E}(V_{0})=\frac{1}{T_{R}}\int_{0}^{T_{R}}E(t)dt,
	\label{time-average_entropy}
\end{equation}
which is plotted as a function of perturbation strength for different particle numbers by solid lines in Fig.~\ref{fig:time-average_entropy}. Apparently, the entanglement measure ${\bar E}$ features a sharp peak at the critical point where the order parameter ${\bar M}$ distinctly changed from $\bar{M}>0$ (dynamical magnetized phase) to $\bar{M}=0$ (dynamical unmagnetized phase), indicating that one can achieve maximal entanglement in the vicinity of the critical point of DPT. We also notice that the entanglement peak at DPT is less sharp with increasing particle number $N$. This is because when we define the entanglement measure $E$, the maximal entanglement $E_{\rm max}$ in the denominator increases logarithmically with $N$. Thus, when $N$ goes larger, the normalization leads to a flatter peak of $E$, making the usage of entropic entanglement measure as an indicator of DPT rather ambiguous. 

\section{Correlation-based entanglement measure for two-mode system}\label{Sec:HZ}

Although the entropy of entanglement is a useful measure to characterize the DPT, it is an entanglement measure only for pure states and therefore cannot account for the effects of finite temperatures. In addition, measuring entropic entanglement requires reconstruction of the quantum states via tomography, which demands the state-of-art technique in experiments even for special systems~\cite{islam2015measuring} and remains challenging for general quantum systems. To circumvent these difficulties, in the following we use an experimentally feasible correlation-based entanglement criterion to detect entanglement undergoing a DPT and discuss the influence of thermal excitations. 

A sufficient entanglement criterion for a two-mode system is the operator product measure $|\langle ab^{\dagger}\rangle|^2>\langle a^\dagger ab^\dagger b\rangle$ given by Hillery and Zubairy (HZ)~\cite{hillery2006entanglement}, where $a$ and $b$ denote the annihilation operators of the two modes. 
For double-well BEC systems, a spin version of HZ criterion has also been developed~\cite{he2011einstein, he2012einstein}. Specifically, the state is entangled if
\begin{equation}
	E_{\rm HZ}=\frac{\Delta^2 J_x+\Delta^2 J_y}{\langle \hat{N} \rangle/2}<1,
	\label{HZ_J}
\end{equation}
where the spin operators $J_{x}=(\psi_{R}^{\dagger}\psi_{L}+\psi_{R}\psi_{L}^{\dagger})/2$, $J_{y}=(\psi_{R}^{\dagger}\psi_{L}-\psi_{R}\psi_{L}^{\dagger})/(2i)$, $J_{z}=(\psi_{R}^{\dagger}\psi_{R}-\psi_{L}^{\dagger}\psi_{L})/2$ with canonical commutation relations $[J_{x},J_{y}]=iJ_{z}$ (and cyclic permutations). The variance of measurements of $J_{x,y}$ is defined as $\Delta^2 J_{x,y}\equiv \langle J^2_{x,y}\rangle-\langle J_{x,y}\rangle^2$, and the expectation values for $\hat{N}=\psi_{R}^{\dagger}\psi_{R}+\psi_{L}^{\dagger}\psi_{L}$ are fixed at $N$ for state~(\ref{two-mode_wave_function}). It is convenient to quantify entanglement using spin-operator methods, and this type of spin-operator variance has been measured experimentally by expanding the two condensates and measuring the absorption imaging average fringe visibility~\cite{esteve2008squeezing, ji2014experimental}. 
We emphasize that the variances $\Delta^2 J_{x,y}$ for the state~(\ref{two-mode_wave_function}) are proportional to the number of atoms, so that the value of $E_{\rm HZ}$ is independent of $N$ as shown in spatial double-well BEC systems~\cite{he2012einstein}.

To get the evolution of the entanglement parameter $E_{\rm HZ}$, we consider the Heisenberg equation for $\langle \psi_{R}^{\dagger}\psi_{R} \rangle$, $\langle \psi_{R}^{\dagger}\psi_{L} \rangle$, $\langle\psi_{L}^{\dagger}\psi_{L}\rangle$, $\langle \psi_{R}^{\dagger}\psi_{R}\psi_{L}^{\dagger}\psi_{L} \rangle$, etc. Due to the nonlinear interaction terms in the Hamiltonian, the sets of equations can't be closed. Thus we truncate the set of equations at the fourth order of operators with mean field approximation, leading to
\begin{widetext}
\allowdisplaybreaks
\begin{eqnarray}
	\frac{d}{dt}\langle\psi_{R}^{\dagger}\psi_{R}\rangle &=& -iV_{p}\left[ \langle\psi_{R}^{\dagger}\psi_{L}\rangle-\langle\psi_{L}^{\dagger}\psi_{R}\rangle \right] , \nonumber\\
	\frac{d}{dt}\langle\psi_{L}^{\dagger}\psi_{L}\rangle &=& -iV_{p}\left[ \langle\psi_{L}^{\dagger}\psi_{R}\rangle-\langle\psi_{R}^{\dagger}\psi_{L}\rangle \right] , \nonumber\\
	\frac{d}{dt}\langle\psi_{R}^{\dagger}\psi_{L}\rangle &=& -i\left[ V_{p}( \langle\psi_{R}^{\dagger}\psi_{R}\rangle-\langle\psi_{L}^{\dagger}\psi_{L}\rangle )+2(E_{s}-E_{m})( \langle \psi_{R}^{\dagger}\psi_{R}^{\dagger}\psi_{R}\psi_{L} \rangle - \langle \psi_{R}^{\dagger}\psi_{L}^{\dagger}\psi_{L}\psi_{L}\rangle ) \right] , \nonumber\\
	\frac{d}{dt}\langle\psi_{R}^{\dagger}\psi_{R}\psi_{L}^{\dagger}\psi_{L}\rangle &\approx& -iV_{p}\left[ \langle \psi_{R}^{\dagger}\psi_{R}\psi_{R}\psi_{L}^{\dagger}\rangle+\langle\psi_{R}^{\dagger}\psi_{L}^{\dagger}\psi_{L}\psi_{L}\rangle-\langle\psi_{R}^{\dagger}\psi_{R}^{\dagger}\psi_{R}\psi_{L}\rangle-\langle\psi_{R}\psi_{L}^{\dagger}\psi_{L}^{\dagger}\psi_{L}\rangle \right] , \nonumber\\
	\frac{d}{dt}\langle\psi_{R}^{\dagger}\psi_{R}^{\dagger}\psi_{R}\psi_{L}\rangle &\approx& -i\left[ V_{p}( \langle\psi_{R}^{\dagger}\psi_{R}^{\dagger}\psi_{R}\psi_{R}\rangle+\langle \psi_{R}^{\dagger}\psi_{R}^{\dagger}\psi_{L}\psi_{L}\rangle -2\langle\psi_{R}^{\dagger}\psi_{R}\psi_{L}^{\dagger}\psi_{L}\rangle)\right.\nonumber\\
	&&\ \ \ \ \ \  \left.+2(E_{s}-E_{m})\langle\psi_{R}^{\dagger}\psi_{R}^{\dagger}\psi_{R}\psi_{L}\rangle( \langle\psi_{R}^{\dagger}\psi_{R}\rangle-\langle\psi_{L}^{\dagger}\psi_{L}\rangle+1) \right] , \nonumber\\
	\frac{d}{dt}\langle\psi_{R}^{\dagger}\psi_{R}^{\dagger}\psi_{L}\psi_{L}\rangle &\approx& -i\left[ 2V_{p}( \langle\psi_{R}^{\dagger}\psi_{R}^{\dagger}\psi_{R}\psi_{L}\rangle-\langle\psi_{R}^{\dagger}\psi_{L}^{\dagger}\psi_{L}\psi_{L}\rangle )+4(E_{s}-E_{m})\langle\psi_{R}^{\dagger}\psi_{R}^{\dagger}\psi_{L}\psi_{L}\rangle( \langle\psi_{R}^{\dagger}\psi_{R}\rangle-\langle\psi_{L}^{\dagger}\psi_{L}\rangle ) \right] ,\nonumber\\
      \frac{d}{dt}\langle\psi_{R}^{\dagger}\psi_{R}^{\dagger}\psi_{R}\psi_{R}\rangle&=&-2iV_{p}\left[ \langle\psi_{R}^{\dagger}\psi_{R}^{\dagger}\psi_{R}\psi_{L}\rangle-\langle\psi_{R}^{\dagger}\psi_{R}\psi_{R}\psi_{L}^{\dagger}\rangle \right] .
\end{eqnarray}
\allowdisplaybreaks[0]
\end{widetext}
Other terms can be derived directly by considering symmetry and conjugate properties of the equations.
\begin{figure*}[t]
	\centering
	\includegraphics[width=0.245\textwidth]{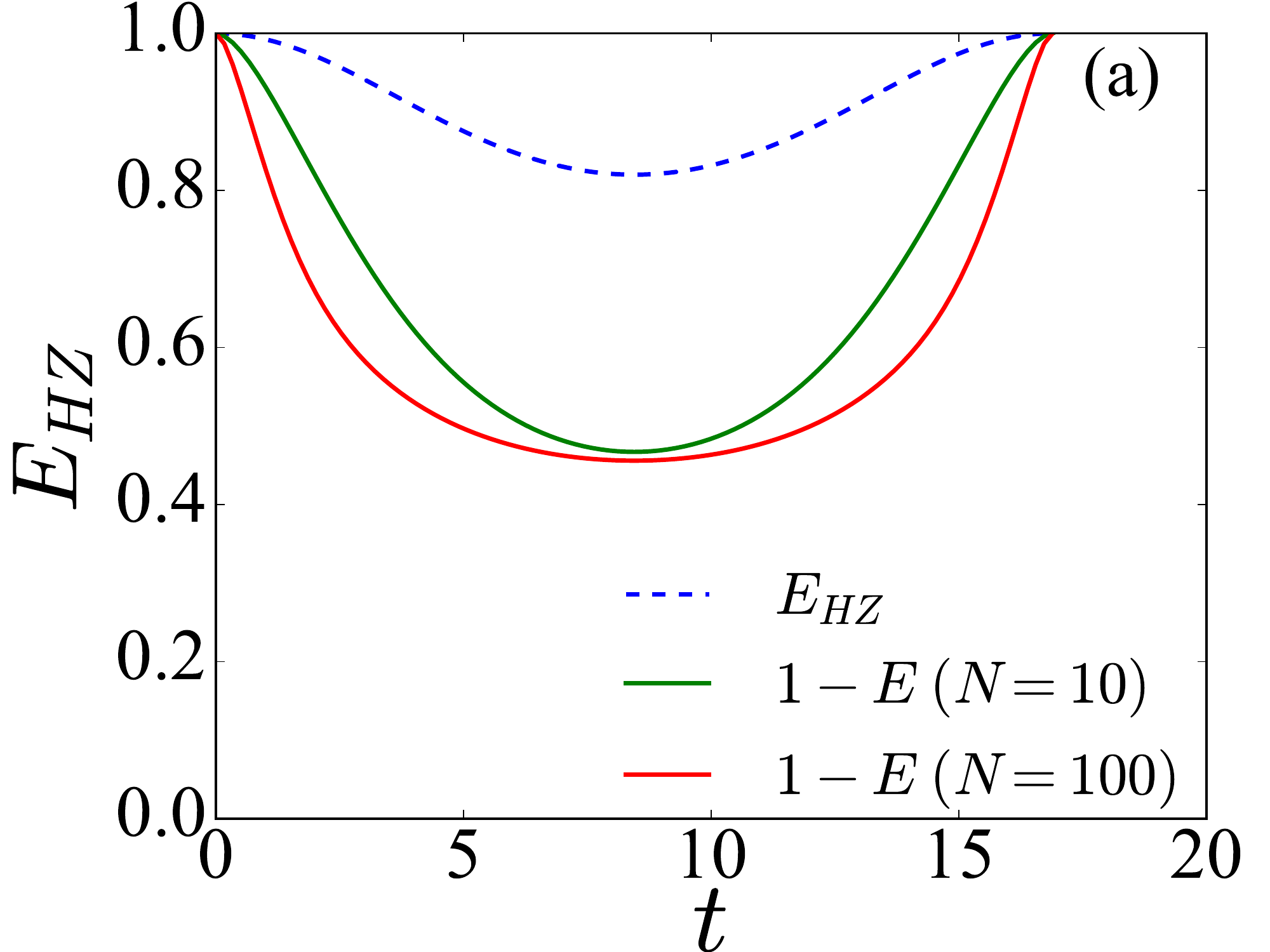}
	\includegraphics[width=0.245\textwidth]{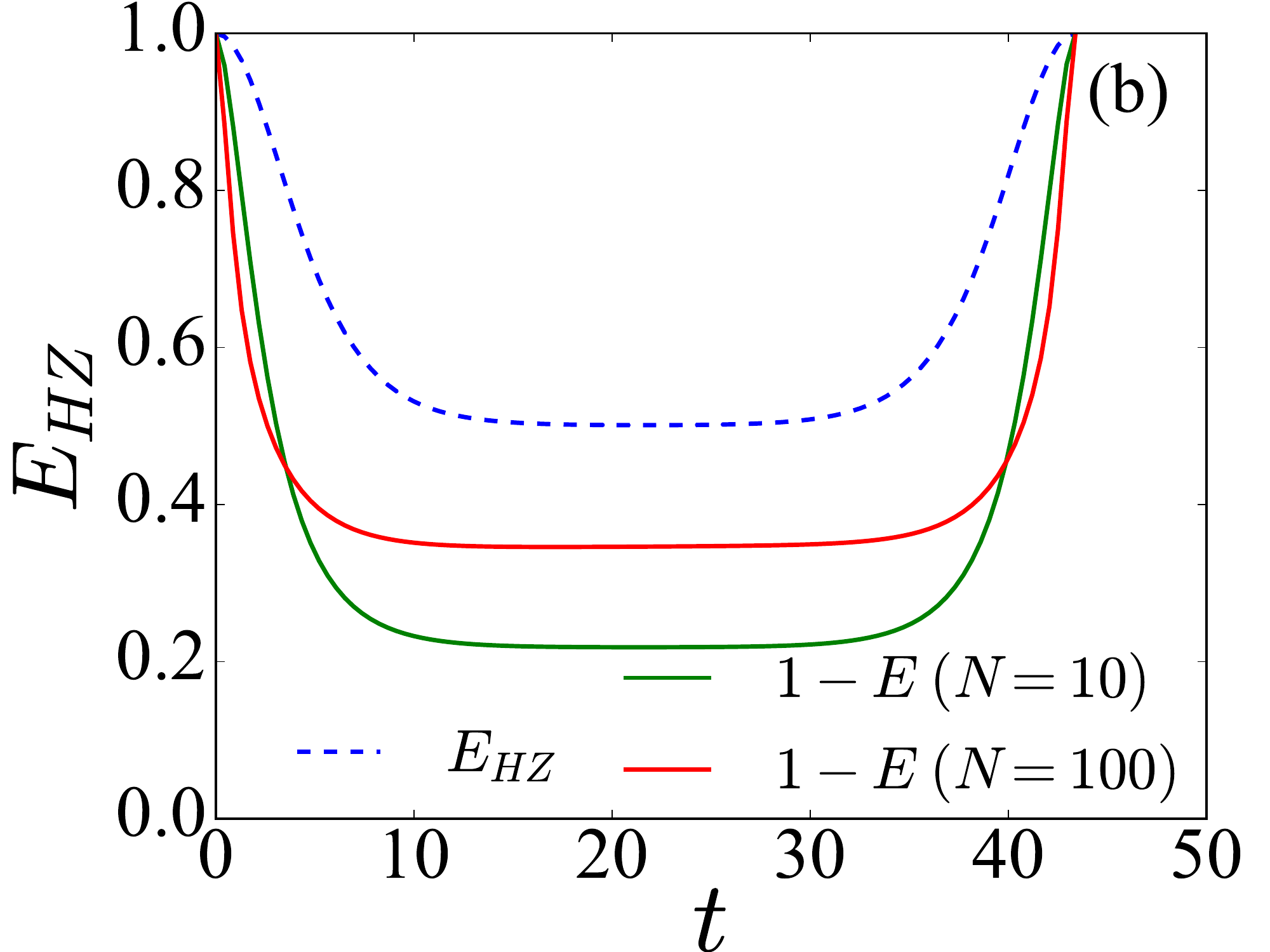}
	\includegraphics[width=0.245\textwidth]{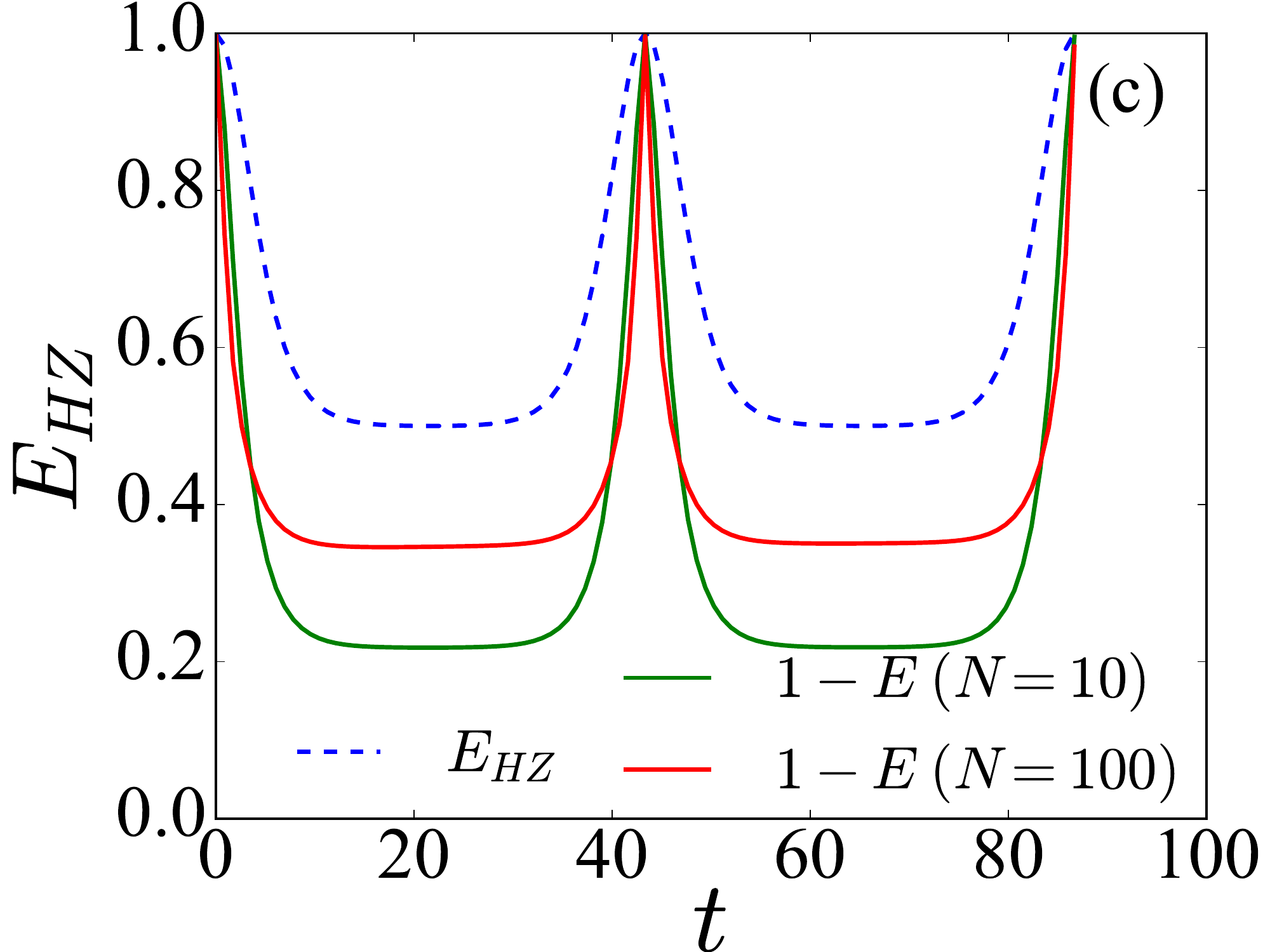}
	\includegraphics[width=0.245\textwidth]{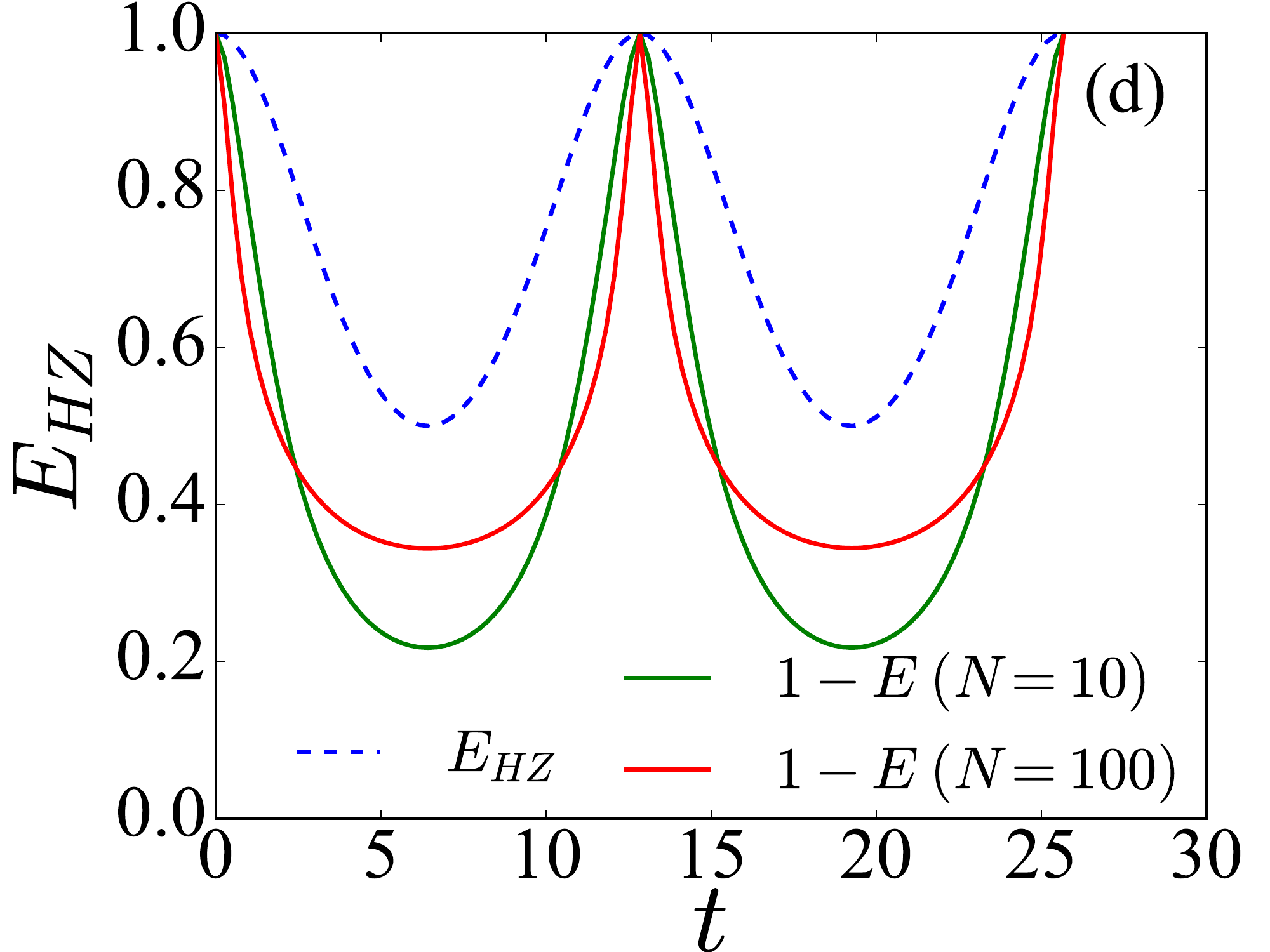}
	\caption{A comparison between the evolution of HZ entanglement parameter $E_{\rm HZ}$ (blue dashed) and the entropic entanglement $1-E$ with total number of atoms $N=10$ (green solid) and $N=100$ (red solid). The strength of external perturbation is (a) $V_{0}=0.6V_{0,{\rm crit}}$, (b) $0.999V_{0,{\rm crit}}$, (c) $1.001V_{0,{\rm crit}}$, and (d) $1.4V_{0,{\rm crit}}$, respectively. The HZ entanglement parameters behaves similarly to the entropic entanglement for different $N$. Other parameters are used as those in Fig.~\ref{fig:entropy}.}
	\label{fig:HZ-compare}
\end{figure*}

By solving the equations above numerically, we show in Fig.~\ref{fig:HZ-compare} the evolution of entropic and HZ entanglement signature. To make a direct comparison, the entropic entanglement measure is plotted as $1-E < 1$.
We find that the evolution of HZ entanglement parameter exhibits the very same qualitative behavior as that of the entropic entanglement measure. Specifically, if the system is in a magnetized state where $|\alpha|=0,\, 1$ and $\langle s_z\rangle\rightarrow \pm 1$, the HZ entanglement parameter $E_{\rm HZ}\rightarrow 1$ showing zero entanglement. On the other hand, the best entanglement would be obtained when the system is in a stripe state where atoms are distributed equally in the two modes with $|\alpha|^2=0.5$ and $\langle s_z\rangle\rightarrow 0$.  

\begin{figure}[t]
	\centering
	\includegraphics[width=0.4\textwidth]{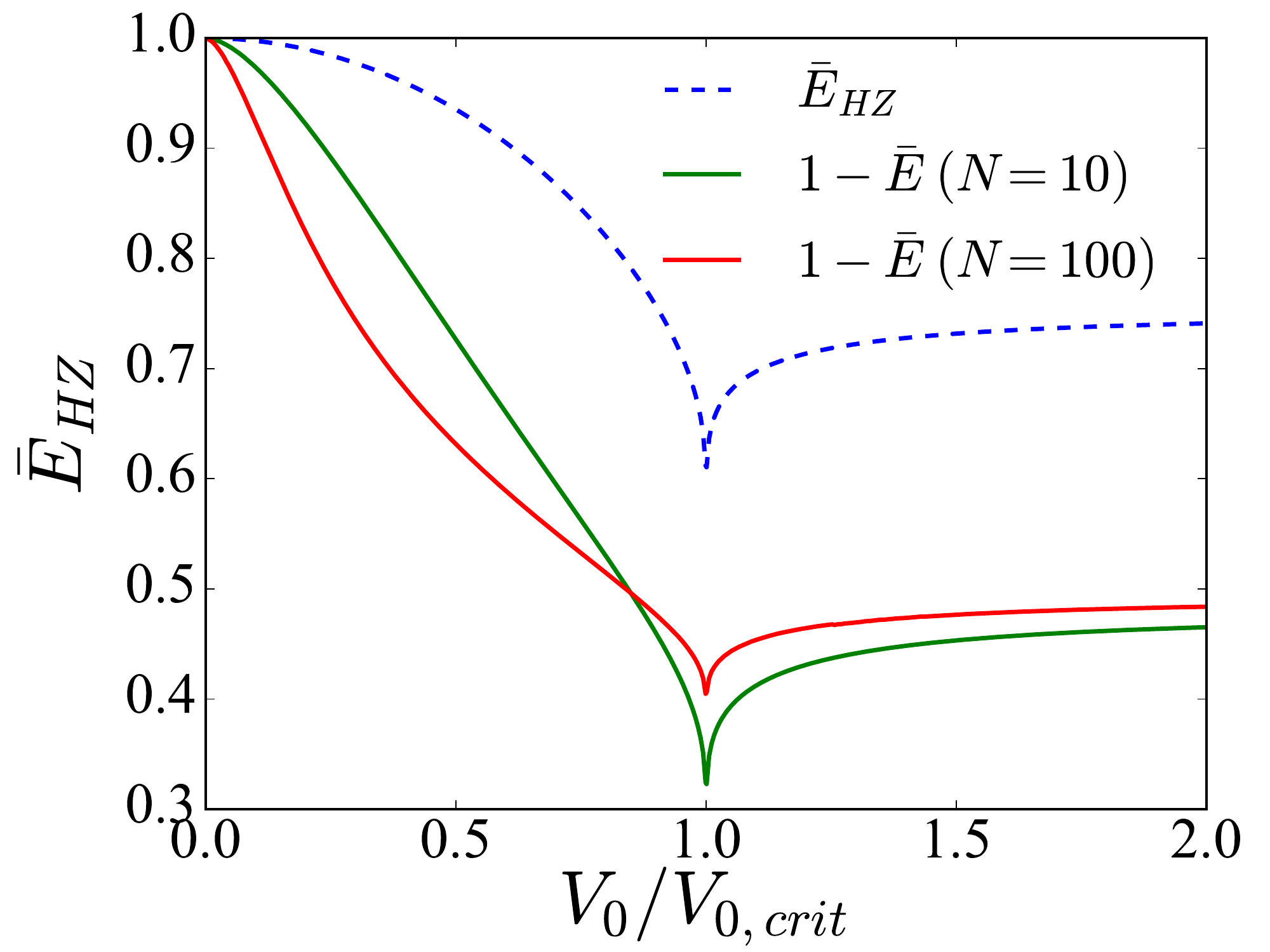}
	\caption{The time-averaged HZ entanglement parameter $\bar{E}_{\rm HZ}$ (blue dashed) calculated as a function of perturbation strength $V_{0}/V_{0,{\rm crit}}$, in comparison with the time-averaged entropic entanglement measure $1-\bar{E}$ with $N=10$ (green solid) and $N=100$ (red solid). In the vicinity of the critical point $V_{0,{\rm crit}}$, $\bar{E}_{\rm HZ}$ shows a discontinuity in the first-order derivative, which signifies the occurrence of DPT. Other parameters are used as those in Fig.~\ref{fig:entropy}.}
	\label{fig:HJ-mean}
\end{figure}

In order to characterize the DPT, we introduce the time-averaged HZ entanglement parameter
\begin{equation}
	\bar{E}_{\rm HZ}=\frac{1}{T_{R}}\int_{0}^{T_R} E_{\rm HZ}(t)dt .
	\label{HZ-t}
\end{equation}
Our results for the entropic entanglement measure $1-\bar{E}$ and the correlation-based HZ entanglement parameter $\bar{E}_{\rm HZ}$ are depicted in Fig.~\ref{fig:HJ-mean}, showing that the HZ measure is an excellent proxy for entropic entanglement measure. The time-averaged HZ entanglement parameter presents a sharp dip in the vicinity of the critical point, which can be used as an indicator for the DPT. Comparing to the entropic measure discussed in Sec.~\ref{Sec:entropy}, the HZ parameter benefits not only from the experimental feasibility, but also from the fact that the dip is not affected by the total numbers of atoms, and thus can identify the DPT in systems with large particle numbers.

\section{Thermal Effects}\label{Sec:thermal}

So far we have studied how the two-mode entanglement can characterize the DPT at zero temperature. In the practical experiments, there are inevitable thermal excitations due to the finite temperature. In general, thermal excitations reduce entanglement because they will cause decoherence and degrade the purity. In the present system, an increasing finite temperature reduces the condensate density $n_{C}$ and consequently changes the interaction energies of the condensate $g_sn_{C}$, as well as the critical perturbation strength $V_{0,{\rm crit}}$. This may change the dynamics of pseudospin and induce new critical points for DPT. We explore how these new critical points at finite temperatures connect with the two-mode entanglement examined by the HZ entanglement criterion. 

In order to study the effects of thermal excitations, we solve the quasiparticle spectra by using Hartree-Fock-Bogoliubov theory with Popov approximation~\cite{pethick2002bose,liao2014spin,chen2015collective,PhysRevA.96.013625}. The general wave function $\Psi_{\pm k_{m},\sigma}$ with $\sigma=\uparrow, \downarrow$ is given by
\begin{equation}
	\Psi_{\pm k_{m},\sigma}(x,t)=e^{-i\mu t\pm ik_{m}x}\left[ \Phi_{\pm k_{m},\sigma}+\delta\Phi_{\pm k_{m},\sigma}(x,t) \right] ,
	\label{general_wave_function}
\end{equation}
where $\mu$ is the chemical potential, $\Phi_{\pm k_{m}, \sigma}$ are the condensate wave functions, and $\delta\Phi_{\pm k_{m},\sigma}(x,t)$ are the fluctuations with the following form
\begin{eqnarray}
	\delta\Phi_{\pm k_{m},\sigma}&=&\psi_{\pm k_{m}+q,\sigma}e^{-i\omega t}+\phi_{\pm k_{m}-q,\sigma}^{\dagger}e^{i\omega t}.
\end{eqnarray}
Here, $q$ is the quasi momentum and $\omega$ is the frequency. It is convenient to solve the Bogoliubov spectrum by expanding $\psi_{\pm k_{m}+q,\sigma}$ and $\phi_{\pm k_{m}-q,\sigma}$ in the Bloch form~\cite{li2013superstripes,poon2016quantum} with basis $\psi_{\tilde{q}+2 \ell k_{m},\sigma}$ and $\phi_{-\tilde{q}+2 \ell k_{m},\sigma}$, 
\begin{eqnarray}
	\psi_{\pm k_{m}+q,\sigma}&=&\sum_{\ell}\psi_{\tilde{q}+2 \ell k_{m},\sigma} ,\nonumber\\
	\phi_{\pm k_{m}-q,\sigma}&=&\sum_{\ell}\phi_{-\tilde{q}+2 \ell k_{m},\sigma} ,
\end{eqnarray}
where $\ell$ is an integer and $|\tilde{q}|<k_{m}$.

Solving the Heisenberg equations $i\partial_{t}\Psi_{\pm k_{m},\sigma}=[\Psi_{\pm k_{m},\sigma}, H]$ by substituting the general wave function of Eq.~(\ref{general_wave_function}) with the mean-field decoupling of interaction and ignoring the anomalous densities $n_{a}=\langle \delta\Phi_{\pm k_m,\sigma}\delta\Phi_{\pm k_m, \sigma'} \rangle$ (i.e. the Hartee-Fock-Bogliubov-Popov approximation)~\cite{pethick2002bose,liao2014spin,chen2015collective,PhysRevA.96.013625}, we can obtain the structure of the energy bands and the density of states $g(\alpha,E)$ of the Bogoliubov quasiparticles at energy $E$ as functions of the superposition coefficient $\alpha$. Please note that we have applied the approximation to the condensate wave functions
\begin{eqnarray}
	\Phi_{k_{m},\uparrow}&=&\sqrt{n_C}\alpha\cos\theta_{k_{m}} , \Phi_{k_{m},\downarrow}=-\sqrt{n_C}\alpha\sin\theta_{k_{m}} , \nonumber\\
	\Phi_{-k_{m},\uparrow}&=&\sqrt{n_C}\beta\sin\theta_{k_{m}} , \Phi_{-k_{m},\downarrow}=-\sqrt{n_C}\beta\cos\theta_{k_{m}} .  \nonumber\\
\end{eqnarray}
The density of excited atoms $n_{\rm ex}$ can be extracted as
\begin{equation}
	n_{\rm ex}=\sum_{s=\pm}\sum_{\sigma=\uparrow,\downarrow}\langle \delta\Psi^{\dagger}_{sk_m,\sigma}\delta\Psi_{sk_m,\sigma} \rangle .
	\label{nex}
\end{equation}
The population of the excited states can then be directly derived as $\Gamma(\alpha,T)={n_{\rm ex}}/{n}$ with $n$ the total density. Using realistic experimental parameters, we estimate that the ratio $\Gamma$ is usually below $0.20$ with $T<70$nK, such that the thermal fluctuations are less important when the system evolves between the magnetized states and stripe state~\cite{poon2016quantum}. 

Because the excitations only take a very small fraction, we expect that the thermal effects on DPT and two-mode entanglement are also limited. Considering the fact that the system becomes a mixed state due to the presence of thermal excitations, the entropic entanglement measure cannot be used to characterize the DPT, while the HZ entanglement parameter $E_{\rm HZ}$ is still suitable within two-mode approximation.

In the low excitation region, we assume that the mixed state of the system takes the following form
\begin{equation}
	\rho=[1-\Gamma(\alpha,T)]\rho_{g}+\Gamma(\alpha,T)\delta\rho ,
	\label{mixed_state}
\end{equation}
where $\rho_{g}$ and $\delta\rho$ are the density operators of ground state and excited states of the system, respectively. Note that the effect induced by all excited states are summed over and denoted by $\delta\rho$, which can be taken as a perturbation when the thermal excitations are only factional at low temperature~\cite{poon2016quantum}. The expectation value of an arbitrary operator $O$ can be calculated by $\langle O\rangle={\rm Tr}(O\rho)$. Thus we can calculate the evolution of the order parameter $\bar{M}$ of DPT and the time-averaged entanglement parameter $\bar{E}_{\rm HZ}$ with thermal excitations at finite temperatures. 
\begin{figure}[t]
	\centering
	\includegraphics[width=0.4\textwidth]{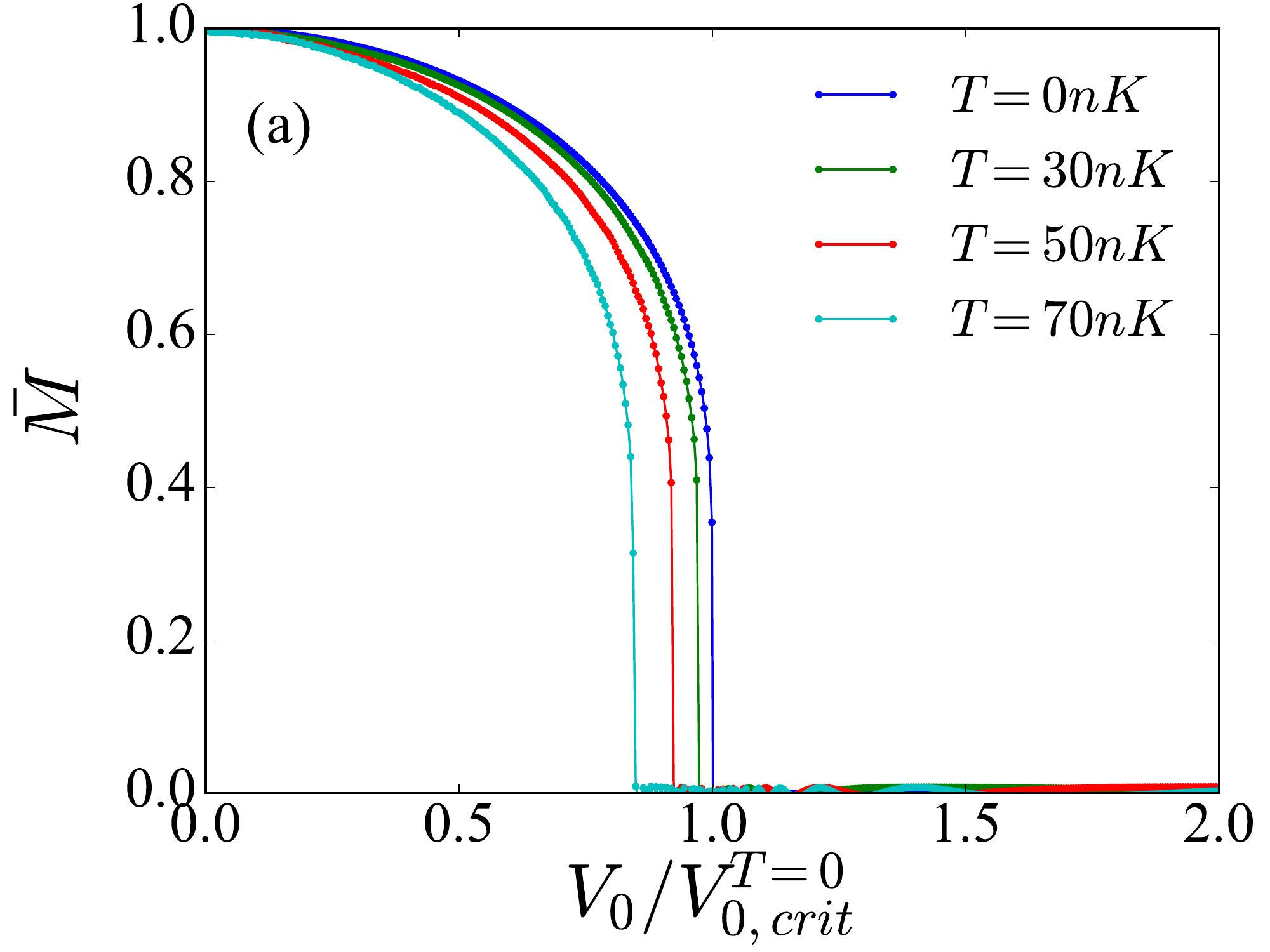}
	\includegraphics[width=0.4\textwidth]{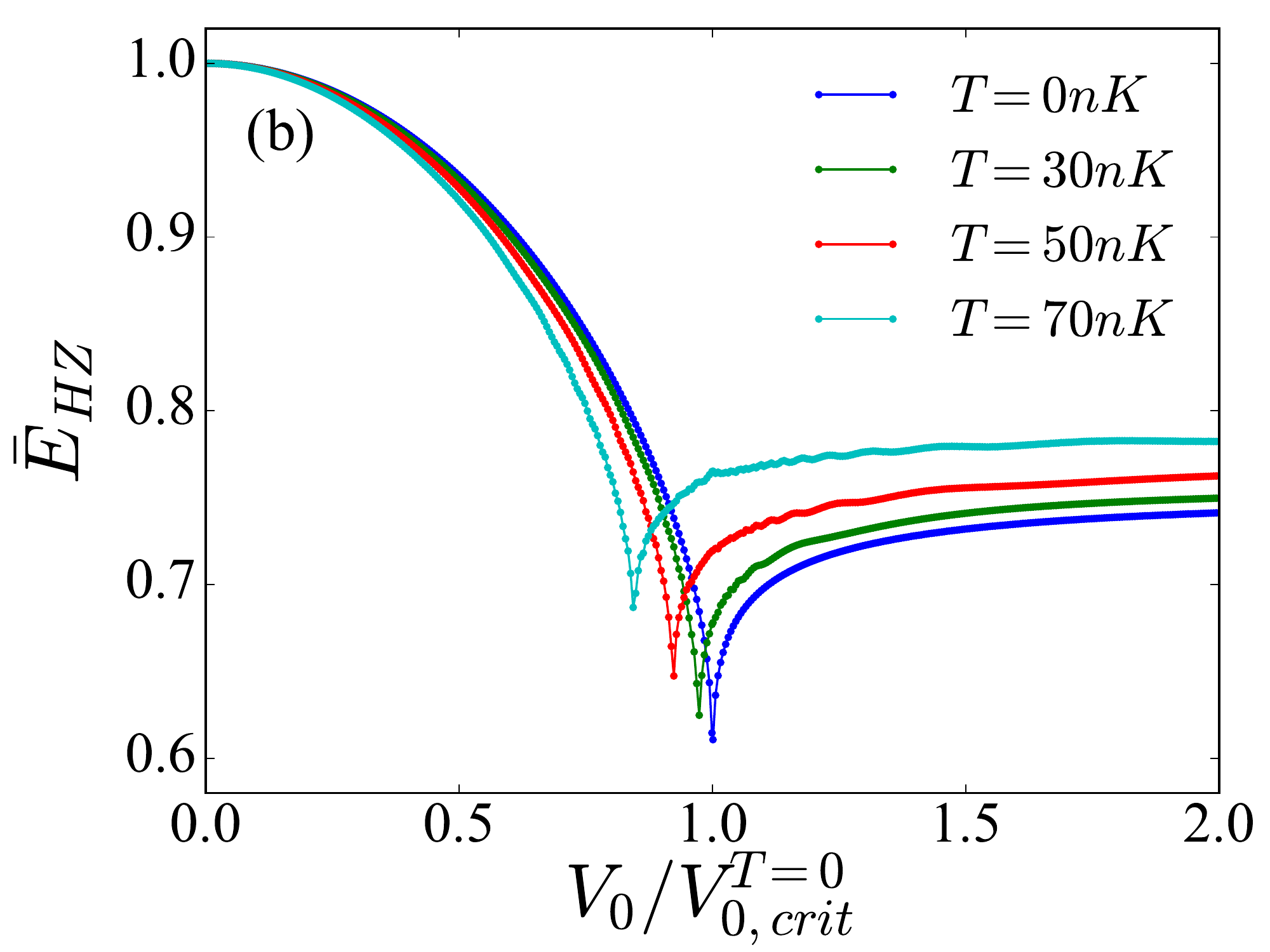}
	\caption{(a) The order parameter $\bar{M}$ and (b) the time-averaged HZ entanglement parameter $\bar{E}_{\rm HZ}$ at temperatures of $T=0$nK (blue), $30$nK (green), $50$nK (red), and $70$nK (cyan) by using Hartree-Fock-Bogoliubov-Popov approximation. The critical point of the DPT shifts towards a smaller value of perturbation strength with elevated temperature, where the HZ entanglement parameter features a sharp dip as an unambiguous signature. Other parameters are used as those in Fig.~\ref{fig:entropy}.}
	\label{fig:EHZ_mean}
\end{figure}

The numerical results of the order parameter $\bar{M}$ are given in Fig.~\ref{fig:EHZ_mean}(a) for different finite temperatures. It can be seen that the transition point shifts towards lower value of $V_0$ with increasing temperature. This is because the thermal excitations reduce the condensate density, which leads to smaller interaction energy and thus induces smaller value of critical perturbation strength defined in Eq.~(\ref{V0c}). The transition points can also be identified from the results of entanglement parameter $\bar{E}_{\rm HZ}$ as shown in Fig.~\ref{fig:EHZ_mean}(b), where the maximal value of entanglement (i.e., minimum of parameter $\bar{E}_{\rm HZ}$ represented by the sharp dip) appear exactly at the transition points where the order parameter $\bar{M}>0$ switching to $\bar{M}=0$ at different temperatures. This observation suggests that the DPT can be characterized by the time-average two-mode entanglement even under finite temperatures. 

We also notice from Fig.~\ref{fig:EHZ_mean}(b) that at the sharp dip, the lowest value of the parameter $\bar{E}_{\rm HZ}$ increases with elevated temperature, indicating that thermal effect is in general detrimental to entanglement. However, since the transition point also shifts with temperature, for a fixed perturbation strength $V_0 < V^{T=0}_{0,{\rm crit}}$ one may find that a lower value of $\bar{E}_{\rm HZ}$ can be obtained with increasing temperature. Physically, this corresponds to the fact that although a finite temperature will inevitably introduce thermal fluctuations and hence degrade entanglement, in certain circumstances it can also bring the system closer to a phase transition point, which features maximal entanglement.

\section{Conclusion}\label{Sec:conclusion}

In summary, we have studied the possibility that using two-mode entanglement in the synthetic (i.e., spin) space to characterize the dynamical phase transitions in a BEC with spin-orbit coupling. By adding an additional lattice potential in the system Hamiltonian as perturbation, we show that the time-averaged entropic entanglement reaches a maximal value at the critical values of perturbation strength where the system is driven from dynamic magnetized phase to a dynamic nonmagnetized phase. The sharp peak of entropic entanglement measure can be used to identify the existence of the DPT. On the other hand, this provides inspiration for generating the maximal entanglement between two modes. Then, considering the difficulty of measuring entropic entanglement in experiments, we have also examined another correlation-based entanglement criterion which is more feasible for experimental test. Our results shows that the time-averaged HZ entanglement parameter is an excellent substitute for entropic entanglement measure, which can not only determine the existence, but also qualitatively characterize the extent of entanglement. Furthermore, it can also be used to account for the effects of thermal excitations induced by finite temperatures. We find that the thermal effects will change the critical point of the DPT, which can be revealed by the shift of the sharp dips of the time-averaged HZ entanglement parameter. 
This work may broaden the understanding of the connection between quantum correlations and dynamical phase transitions in the SOC systems with interactions. 

In the end, we would briefly comment about the experimental feasibility of the study. The synthetic spin-orbit coupling in ultracold atomic gases for pseudospin-1/2 systems has been realized experimentally~\cite{lin2011spin, wang2012spin, cheuk2012spin, ji2014experimental}. The periodic perturbation $V_{\rm ex}$ is simply a lattice potential which can be generated by standing-wave lasers with wave vector $k_{m}$. The time-averaged two-mode entanglement parameter $\bar{E}_{\rm HZ}$ can be detected by measuring the evolution of pseudospin variance of the BEC which has been realized in many experiments for quantum squeezing and metrology~\cite{esteve2008squeezing,riedel2010atom,luo2017deterministic}. Thus, we expect that the system can be readily prepared and investigated with present experimental technique.

\begin{acknowledgements}

We acknowledge illuminating discussions with X.-J. Liu, T. F. Jeffrey Poon, and H. Zhai. This work is supported by the Ministry of Science and Technology of China (Grant No. 2016YFA0301302), National Natural Science Foundation of China (Grants No. 11622428, No. 11274025, No. 61475006, No. 11434011, No. 11522436, and No. 11774425), the Research Funds of Renmin University of China (Grants No. 10XNL016 and No. 16XNLQ03), and the National 973 Program (Grant No. 2014CB921403).

\end{acknowledgements}

\bibliography{DPTandEntanglement}
\end{document}